\documentclass{article}
\usepackage{lmodern}
\usepackage[T1]{fontenc}
\usepackage[latin9]{inputenc}
\usepackage{geometry}
\usepackage{array}
\usepackage{float}
\usepackage{multirow}
\usepackage{amsmath}
\usepackage{amssymb}
\usepackage{authblk}
\usepackage{setspace}
\usepackage{esint}
\usepackage[authoryear]{natbib}
\PassOptionsToPackage{normalem}{ulem}
\usepackage{ulem}
\usepackage{url}
\usepackage{tikz}
\usepackage{graphicx}
\usepackage{longtable}
\usepackage{url}
\usepackage{amsmath}
\usepackage{longtable}
\usepackage{subfigure}
\usepackage[figuresright]{rotating}
\usepackage{color}
\usepackage{threeparttable}
\usepackage{multirow}   
\usepackage{enumerate}
\usepackage{graphicx}
\usepackage{algorithm}
\usepackage{algorithmicx}
\usepackage{bm}
\usepackage{authblk}
\usepackage[noend]{algpseudocode}
\makeatletter  
\def\BState{\State\hskip-\ALG@thistlm}
\makeatother

\newcommand{\tabincell}[2]{\begin{tabular}{@{}#1@{}}#2\end{tabular}}

\def\bfe{{\mathbf{e}}}

\def\bfx{{\mathbf{x}}}
\def\bfy{{\mathbf{y}}}

\def\bfI{{\mathbf{I}}}

\def\bfP{{\mathbf{P}}}

\def\bfX{{\mathbf{X}}}

\def\pr{\mathrm{Pr}}

\def \bftheta {\boldsymbol{\theta}}

\def \bfbeta {\boldsymbol{\beta}}
\def \bfgamma {\boldsymbol{\gamma}}
\def \bfGamma {\boldsymbol{\Gamma}}

\def \bfbeta {\boldsymbol{\beta}}

\usetikzlibrary{
  shapes.multipart,
  matrix,
  positioning,
  shapes.callouts,
  shapes.arrows,
  calc}

 \usetikzlibrary{external}

 \newcommand{\content}[2]{%
    \let\mymatrixcontent\empty
    \foreach \j in {1,...,#2}{
        \foreach \i [evaluate=\i as \myindex using {int((\j-1)*#1+\i)}] in {1,...,#1}{%
            \begingroup\edef\x{\endgroup
                \noexpand\gappto\noexpand\mymatrixcontent{\myindex\&}}\x
            }%
        \gappto\mymatrixcontent{\\}%
    }
}

\tikzset{
  mtx/.style={
    matrix of math nodes,
    left delimiter={[}, right delimiter={]}
  },
  hlt/.style={opacity=0.1, line width=4 mm, line cap=round},
  hltr/.style={opacity=0.5, rounded corners=2pt, inner sep=-1pt}
}

\usetikzlibrary{shapes, arrows, calc, positioning,matrix}

\usetikzlibrary{fit,positioning}

\usetikzlibrary{bayesnet}

\usetikzlibrary{matrix}
\usetikzlibrary{decorations.pathreplacing}

\tikzset{
data/.style={circle, draw, text centered, minimum height=3em ,minimum width = .5em, inner sep = 2pt},
empty/.style={circle, text centered, minimum height=3em ,minimum width = .5em, inner sep = 2pt},
}

\begin{document}

\title{Joint Analysis of Individual-level and Summary-level GWAS Data by Leveraging Pleiotropy}


\author[1,2]{Mingwei Dai}
\author[3]{Xiang Wan}
\author[4]{Hao Peng}
\author[1]{Yao Wang}
\author[5]{Yue Liu}
\author[6, *]{Jin Liu}
\author[1, *]{Zongben Xu}
\author[2, *]{Can Yang} 
\affil[1]{School of Mathematics and Statistics, Xi'an Jiaotong University, Xi'an, China}
\affil[2]{Department of Mathematics, Hong Kong University of Science and Technology, Hong Kong}
\affil[3]{ShenZhen Research Institute of Big Data, ShenZhen, China}
\affil[4]{School of Business Administration, Southwestern University of Finance and Economics, Chengdu, China}
\affil[5]{Xiyuan Hospital of China Academy of Chinese Medical Sciences, Beijing, China}
\affil[6]{Centre for Quantitative Medicine, Duke-NUS Medical School, Singapore}

\renewcommand\Authands{ and }

\maketitle

\begin{abstract}
A large number of recent genome-wide association studies (GWASs) for complex phenotypes confirm the early conjecture for polygenicity, suggesting the presence of large number of variants with only tiny or moderate effects. However, due to the limited sample size of a single GWAS, many associated genetic variants are too weak to achieve the genome-wide significance. These undiscovered variants further limit the prediction capability of GWAS. Restricted access to the individual-level data and the increasing availability of the published GWAS results motivate the development of methods integrating both the individual-level and summary-level data. How to build the connection between the individual-level and summary-level data determines the efficiency of using the existing abundant summary-level resources with limited individual-level data, and this issue inspires more efforts in the existing area. 

In this study, we propose a novel statistical approach, LEP, which provides a novel way of modeling the connection between the individual-level data and summary-level data. LEP integrates both types of data by \underline{LE}veraing \underline{P}leiotropy to increase the statistical power of risk variants identification and the accuracy of risk prediction. The algorithm for parameter estimation is developed to handle genome-wide-scale data. Through comprehensive simulation studies, we demonstrated the advantages of LEP over the existing methods. We further applied LEP to perform integrative analysis of Crohn's disease from WTCCC and summary statistics from GWAS of some other diseases, such as Type 1 diabetes, Ulcerative colitis and Primary biliary cirrhosis. LEP was able to significantly increase the statistical power of identifying risk variants and improve the risk prediction accuracy from 63.39\% ($\pm$ 0.58\%) to 68.33\% ($\pm$ 0.32\%) using about 195,000 variants.

The LEP software is available at 
\url{https://github.com/daviddaigithub/LEP}.


\end{abstract}

\section{Introduction}
The explosive growth of genome-wide association studies (GWASs) is leading to a new era for understanding the genetic underpinnings of complex phenotypes. As of March 2018, more than 3,300 GWASs have been conducted for hundreds of complex traits, about 59,000 statistically significant (at the genome-wide $p$-value threshold of $5\times 10^{-8}$) associations between genetic variants and the complex traits have been reported \citep{welter2014nhgri}. However, most of these variants contribute relatively small increments of risk and only explain a small portion of variation in complex diseases. Recently, many GWASs have suggested that complex diseases are affected by the collective effects of many genetic variants, each of which may have only a small effect, known as `polygenicity' \citep{visscher201710,yang2016introduction}. Due to the limited sample size of a single GWAS, many of the individual effects of genetic variants could not achieve the genome-wide significance and thus remain undiscovered. 

Although the available experimental sample sizes have dramatically increased in many recent studies, a comprehensive integration over the existing data resources may further advance our understandings of complex traits \citep{flannick2016type,khera2017genetics}. To this end, many methods have been proposed to improve the analysis by combining datasets from resources of multiple biological platforms. These methods could be divided into three categories according to their inputs of different data types. The methods in the first group (e.g., \citep{li2014improving}) work with related individual-level datasets to achieve better performance by jointly analyzing individual-level data from different sources. As the availability of summary statistics from GWASs increases, the methods in the second group (e.g., \citep{chung2014gpa,liu2016eps,liu2017llr,mak2017polygenic,turley2018multi,zhu2015meta}) working with summary statistics are more convenient to be applied because the summary-level data sets are easily accessible. (see the detailed review of these methods in \citep{pasaniuc2017dissecting}). Compared to the methods in the first group, the methods in the second group have the advantage of computation, but they might sacrifice statistical efficiency. To make the most efficient use of the available resources of GWAS data, the methods in the third group work with both individual-level data and summary-level data simultaneously, e.g., \citep{dai2017igess,purcell2009common,shi2016winner}.

The methods in the third group (e.g., IGESS \citep{dai2017igess}) often make a homogeneous assumption for risk variants, i.e., all risk variants are assumed to be shared in the individual-level data and the summary-level data. This assumption makes sense when the individual-level data and the summary-level data correspond to the same disease and the samples are from the same population. However, such a requirement can be too restricted to allow this group of method to make most efficient use of existing data resources. In fact, GWASs have identified genetic variants that can affect multiple seemly different complex traits. This phenomenon has been known as `Pleiotropy' \citep{stearns2010one}. Accumulating evidence suggests pleiotropy widely exists among complex diseases \citep{yang2015implications,sivakumaran2011abundant}, such as autoimmune diseases \citep{cotsapas2011pervasive} and psychiatric disorders \citep{cross2013identification}. Examples include: 
SNP rs1800693 in the major TNF-receptor gene (\emph{TNFR1}) implicated in multiple sclerosis (MS) has also been associated with ankylosing spondylitis (AS) \citep{visscher201710}; the protein tyrosine phosphatase non-receptor type 22 (\emph{PTPN22}) is associated with rheumatoid arthritis (RA) and systemic lupus erythematosus (SLE) \citep{begovich2004missense}. Therefore, it is necessary to get rid of the homogeneous assumption and effectively integrate individual-level data and summary-level data by leveraging abundant pleiotropic information among complex traits.

In this study, we propose a statistical approach, LEP, to integrate both the individual-level and summary-level data by \underline{LE}veraging \underline{P}leiotropy when the corresponding traits are of a pleiotropic relationship. An efficient variational-inference-based algorithm has been developed such that it can fit the model in an efficient manner. Not only does LEP make efficient use of GWAS data but also provides a new way of characterizing the pleiotropy. Through comprehensive simulation studies, we demonstrated the effectiveness of LEP by leveraging pleiotropy in the presence of heterogeneity among the individual-level and summary-level data, which takes the advantages over the existing methods. We then applied LEP to perform integrative analysis on individual-level data of Crohn's disease from WTCCC and summary statistics of some other diseases. LEP was able to significantly increase the statistical power of identifying risk variants, and correspondingly improve the risk prediction.

\section{LEP}
\subsection{Model}
Given a trait named T, suppose we have an individual-level genotype data set $\bfX \in \mathbb{R}^{N\times M}$ from $N$ individuals and their corresponding phenotypes $\bfy \in \mathbb{R}^N$, where $M$ is the number of SNPs. Without loss of generality, we assume both $\bfX$ and $\bfy$ have been centered. In addition, we collect summary statistics, i.e., $p$-values from $K$ independent GWASs in matrix $\mathbf{P}= [p_{jk}] \in \mathbb{R}^{M\times K}$, where $p_{jk}$ corresponds to the $p$-value of the $j$-th SNP in the $k$-th GWAS. Suppose these $K$ GWASs are for $K$ different traits, each of which shares certain associated variants with trait T. First, we consider the following linear model for the genotype data,
\begin{equation}\label{eq1}
  \bfy = \bfX \bfbeta + \bfe,
\end{equation}
where $\bfbeta= [\beta_1,\dots,\beta_M]^T$ is a vector of effect sizes, and $\bfe$ is the independent random error with the distribution $N(\bfe|\mathbf{0}, \sigma^2_e \bfI)$. The identification of risk variants is equivalent to the variable selection in Eq.~\eqref{eq1}. A binary variable $\gamma_j$ is introduced to indicate whether $\beta_j$ is zero or not. Assuming the 
spike-and-slab prior \citep{mitchell1988bayesian} for $\beta_j$, 
\begin{equation}{\label{eqprior2beta}}
\begin{split}
\beta_j|\gamma_j, \sigma^2_\beta \sim \left\{
\begin{aligned}
&N(\beta_j | 0,\sigma_{\beta}^2) &  &\mathrm{if}~ \gamma_{j} = 1,\\
&\delta_0(\beta_j) &  &\mathrm{if}~\gamma_{j} = 0, \\
\end{aligned}
\right.
\end{split}
\end{equation}
where $\beta_j$ from the slab group follows a Gaussian distribution $N(\beta_j | 0,\sigma_{\beta}^2)$ and $\beta_j$ from the spike group corresponds to the Dirac function centered at zero, and
$\gamma_j$ is assumed to follow the Bernoulli distribution $\mathrm{Bern}(\gamma_j|\pi)$,
\begin{equation}\label{gammaBern}
  \gamma_j|\pi \sim \pi^{\gamma_j}(1-\pi)^{1-\gamma_j}.
\end{equation}

Second, we assume that the $p$-values from the $k$-th GWAS follow a mixture distribution, a binary variable $\Gamma_{jk}$ is introduced to indicate whether the $j$-th SNP is associated with the $k$-th trait,
\begin{equation}{\label{eqpriorpvalue}}
\begin{split}
p_{jk}|\Gamma_{jk}, \alpha_k \sim \left\{
\begin{aligned}
&\mathrm{U}(0,1) &  &\mathrm{if}~\Gamma_{jk} = 0, \\
&\mathrm{Beta}(p_{jk}|\alpha_k, 1) & &\mathrm{if}~ \Gamma_{jk} = 1,\\
\end{aligned}
\right.
\end{split}
\end{equation}
where $p$-values from the null group follow the uniform distribution $\mathrm{U}(0,1)$ and $p$-values of the $k$-th study from the non-null group follow the Beta distribution with parameter $(\alpha_k,1)$.

To model the pleiotropic relationship between trait T and $k$-th trait of summary
statistics, we define, 
\begin{equation}{\label{uv}}
  \begin{split}
    u_k := \pr(\Gamma_{jk} = 1| \gamma_j = 1), \\
    v_k := \pr(\Gamma_{jk} = 0| \gamma_j = 0).
  \end{split}
\end{equation}
Here $u_k$ is the conditional probability that a variant is associated with 
the $k$-th trait given that this variant is associated with trait T, and $v_k$ is defined for the opposite case. The introduction of these two parameters enables the proposed model to adapt to the heterogeneous cases which cover the homogeneous case ($u_k = 1$, $v_k = 1$) as a special case. The pair of parameters characterizes the degree of pleiotropic effects of trait T and the $k$-th trait.

\begin{figure}
\begin{center}
   \begin{tikzpicture}
      \node[const] (pi) {$\pi$} ; %
      \node[latent, right=of pi] (gammap) {$\gamma_j$} ; %
      \node[latent, right=of gammap,xshift=0.6cm,yshift = 0.3cm] (betap) {$\beta_j$} ; %
      \node[const, above=of betap,xshift=0cm,yshift = -0.3cm] (sigmab) {$\sigma^2_{\beta}$} ; %
      \node[latent, right=of gammap,xshift=0.6cm,yshift = -0.6cm] (gammak) {$\Gamma_j^{k}$} ; %
             \node[const, below=of gammak,xshift = 0.0cm, yshift = 0.61cm] (uv) {$(u_k, v_k)$} ; %

      \node[obs, right=of gammak] (pk) {$p_{jk}$} ; %
      
      \node[const, below=of pk,xshift=0.0cm,yshift = 0.5cm] (alphak) {$\alpha_k$} ; %
             \edge {pi} {gammap} ; %
       \edge {gammap} {betap} ; %
       \edge {sigmab} {betap} ; %
              \edge {gammap} {gammak} ; %
              \edge {gammak} {pk} ; %
               \edge {alphak} {pk} ; %
               \edge {uv} {gammak} ;%
      \node[obs, right=of betap, xshift = 1.8cm] (yn) {$y_n$} ; %
        \node[const, above=of yn,xshift=0cm,yshift = -0.3cm] (sigmae) {$\sigma^2_{e}$} ; %
        \edge {sigmae} {yn}; %
      \node[const, below=of yn, yshift = 0.5cm] (xn) {$x_n$} ; %
          \edge {betap} {yn} ; %
        \edge {xn} {yn}  ; %
              \plate[inner sep=0.24cm, xshift=-0.12cm, yshift=-0.1cm] {plate1} {(gammak)(pk)} {\emph{K}}; %
        \plate[inner sep=0.35cm, xshift= 0.06cm, yshift=-0.05cm] {plate2} {(plate1)(gammap) (betap)  (pk)} {\emph{M}}; %
              \plate[inner sep=0.15cm, xshift=-0.12cm, yshift=0.12cm] {plate3} {(yn) (xn)} {\emph{N}}; %

   \end{tikzpicture}
    \caption{Graphical Model of LEP, the circles denote the observed or unobserved randoms variables, the shaded circles denote the observed ones. Other nodes are those fixed variables or parameters.}
   \label{fig:graphical_lep}
   \end{center}
\end{figure}
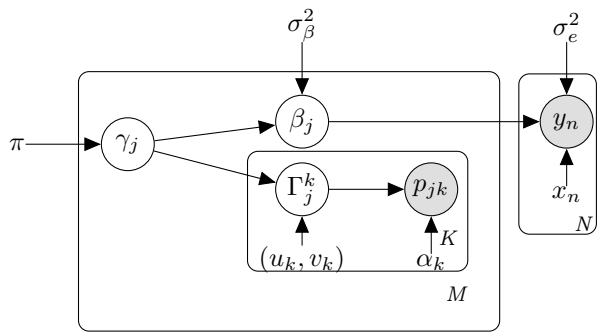

Furthermore, we assume $\Gamma_{1j},\hdots,\Gamma_{jK}$ are conditional independent given $\gamma_j$, denote $\bfGamma = [\Gamma_{jk}] \in \mathbb{R}^{M \times K}$. Then we have
\begin{equation}{\label{eqidd}}
  \pr(\bfGamma|\bfgamma) = \prod_{j=1}^{M} \prod_{k=1}^{K} \pr( \Gamma_{jk} | \gamma_j).
\end{equation}
Let $\bftheta=\{\pi,\sigma^2_{\beta},\sigma^2_e,\{\alpha_k,u_k, v_k\}^K_{k=1}\}$ be the collection of model parameters. The probabilistic model can be written as
\begin{align}{\label{eqjointd}}
  &\pr(\bfy,\bfP,\bfbeta,\bfGamma,\bfgamma|\bfX; \bftheta) \nonumber\\
  = &\pr(\bfy|\bfX,\bfbeta; \bftheta) \pr(\bfbeta| \bfgamma;\bftheta)\pr(\bfgamma|\bftheta)\pr(\bfP|\bfGamma;\bftheta)\pr(\bfGamma|\bfgamma;\bftheta).
\end{align}
\textcolor{black}{Its graphical representation could be referred in Fig.~\ref{fig:graphical_lep}.}

We aim to obtain $\hat{\bftheta}$ (the estimate of $\bftheta$) by maximizing the marginal likelihood
\begin{align}\label{margloglik}
\pr(\bfy,\bfP|\bfX; \bftheta) = \sum_{\bfgamma,\bfGamma}\int_{\bfbeta}\pr(\bfy,\bfP,\bfbeta,\bfgamma, \bfGamma|\bfX; \bftheta)d \bfbeta,
\end{align}
and then have the posterior of latent variable $\{\bfbeta, \bfgamma, \bfGamma\}$ as $\pr(\bfbeta, \bfgamma,\bfGamma | \bfy,\bfX, \bfP; \hat\bftheta)$. Although the above LEP model was designed for the traits with the Gaussian noise, we may use the logit or probit link function for handling case-control studies. Empirical results (e.g., \citep{dai2017igess}) suggest the performance of generalized linear models does not show significant improvements if the sample size is moderate (e.g., a few thousands) while the computation burden might be more expensive. In fact, the effectiveness of linear models applied for the analysis of case-control GWAS datasets has been justified in \citep{kang2010variance}. Therefore, we developed LEP based on the Gaussian assumption, and demonstrate its effectiveness on the case-control studies by both simulation study and real data analysis.

\subsection{Algorithm}
The intractability of the exact evaluation of Eq.~(\ref{margloglik}) sets up a challenge for solving the model. To address this challenge, we derive an efficient algorithm based on variational inference \citep{bishop2006pattern}. To get rid of the Dirac function, the reparametrization is conducted as follows. Assume $\tilde{\beta}_j$ follows a Gaussian distribution $N(\tilde{\beta}_j|0,\sigma^2_\beta)$ and $\gamma_j$ follows a Bernoulli distribution $\mathrm{Bern}(\gamma_j|\pi)$, respectively. Clearly, their product $\tilde{\beta}_j\gamma_j$ follows the same distribution as $\beta_j$ in Eq.~\eqref{eqprior2beta}. With this reparametrization, the joint model Eq.~\eqref{eqjointd} becomes
\begin{align}{\label{eqjointd2}}
  &\pr(\bfy,\bfP,\tilde\bfbeta,\bfgamma,\bfGamma|\bfX; \bftheta) \nonumber\\
  = &\pr(\bfy|\bfX,\tilde{\bfbeta}, \bfgamma; \bftheta) \pr(\tilde{\bfbeta};\bftheta)\pr(\bfgamma|\bftheta)\pr(\bfP|\bfGamma;\bftheta)\pr(\bfGamma|\bfgamma;\bftheta),
\end{align}
where
\begin{align}
\pr(\bfy|\bfX,\tilde{\bfbeta}, \bfgamma; \bftheta) &= N(\bfy| \sum_j \bfx_j \tilde{\beta}_j \gamma_j, \sigma^2_e \bfI),\nonumber\\
\pr(\tilde{\bfbeta}|\bftheta) &= \prod^M_j N(\tilde{\beta}_j | 0, \sigma^2_\beta),\nonumber\\
\pr(\bfgamma|\bftheta) &=\prod^M_j \pi^{\gamma_j}(1-\pi)^{1-\gamma_j},\nonumber\\
\pr(\bfP|\bfGamma;\bftheta) &=\prod^M_j \prod^K_k \Big( \alpha_k p^{\alpha_k-1}_{jk}\Big)^{\Gamma_{jk}},\nonumber \\
  \pr(\bfGamma|\bfgamma;\bftheta) &= \prod_{j}^{M} \prod_{k}^{K} \pr( \Gamma_{jk} | \gamma_j). \nonumber
\end{align}

Clearly, the Dirac function is not involved and the prior of $\tilde{\beta}_j$ does not depend on $\gamma_j$ after reparametrization. Next, we aim to find a variational approximation $q(\tilde{\bfbeta},\bfgamma, \bfGamma)$ to the true posterior  $\pr(\tilde{\bfbeta},\bfgamma, \bfGamma|\bfy,\bfX,\bfP; \bftheta)$. Then we will have a lower bound of the logarithm of the marginal likelihood
\begin{equation}\label{lowerbound}
\begin{split}
  &\log \pr(\bfy, \bfP |\bfX; \bftheta) \\
  = &\log \sum_{\bfgamma,\bfGamma} \int_{\tilde\bfbeta} \pr(\bfy, \bfP,\tilde{\bfbeta},\bfgamma,\bfGamma|\bftheta) d \tilde\bfbeta  \\
  \geq &\sum_{\bfgamma, \bfGamma} \int_{\tilde\bfbeta} q(\tilde\bfbeta,\bfgamma, \bfGamma) \log \frac{\pr(\bfy, \bfP,\tilde\bfbeta,\bfgamma, \bfGamma|\bfX;\bftheta)}{q(\tilde\bfbeta,\bfgamma, \bfGamma)} d \tilde\bfbeta   \\
   = &\mathbb{E}_q[\log \pr(\bfy, \bfP,\tilde\bfbeta,\bfgamma,\bfGamma|\bfX;\bftheta) - \log q(\tilde\bfbeta,\bfgamma, \bfGamma)]  \\
  := &\mathcal{L}(q),
\end{split}  
\end{equation}
where the inequality follows Jensen's inequality, and the equality holds if and only if $q(\tilde\bfbeta,\bfgamma,\bfGamma)$ is the true posterior $\pr(\tilde\bfbeta,\bfgamma,\bfGamma |\bfy,\bfP,\bfX;\bftheta)$. Now, we can iteratively maximize $\mathcal{L}(q)$ instead of directly working with the marginal log-likelihood. To make it feasible to evaluate this lower bound, we assume that $q(\tilde\bfbeta,\bfgamma, \bfGamma)$ can be factorized as
\begin{equation*}{\label{eqassump}}
  q(\tilde\bfbeta,\bfgamma, \bfGamma) = \prod_{j=1}^M q_j(\tilde{\beta}_j,{\gamma_j},
  {\bfGamma_j}),
\end{equation*}
where $\bfGamma_j = [\Gamma_{jk}] \in \mathbb{R}^{1 \times K}$. This is the only assumption we made in the variational inference. According to the property of factorized distributions in variational inference \citep{bishop2006pattern}, we can obtain the best approximation as
\begin{equation*}
  \log q_j({\tilde{\beta}_j},{\gamma_j},\bfGamma_j) = \mathbb{E}_{i\neq j}[\log \pr(\bfy, \bfP,\tilde{\bfbeta},\bfgamma,\bfGamma|\bfX; \bftheta)] + \mbox{Const},
\end{equation*}
where the expectation is taken with respect to all of the other factors $\{ q_i({\tilde{\beta}_i},{\gamma_i},\bfGamma_i)\}$ for $i \neq j$.
As $q(\tilde{\beta}_j,\gamma_j,\bfGamma_j)=q(\tilde{\beta}_j|\gamma_j) q(\bfGamma_j|\gamma_j) q(\gamma_j)$, after some derivations (refer to Section 1.1 of Supplementary document), we have
\begin{equation}\label{q_j}
\begin{split}
    q(\tilde{\beta}_j,\gamma_j,\bfGamma_j) = &\left(\pi_j N(\tilde{\beta}_j|\mu_j, s^2_j) \prod_{k=1}^K q(\Gamma_{jk}|\gamma_j = 1) \right)^{\gamma_j} \\
    &\left((1-\pi_j)N(\tilde{\beta}_j|0,\sigma^2_{\beta})  \prod_{k=1}^K q(\Gamma_{jk}|\gamma_j = 0) \right)^{1-\gamma_j},
\end{split}
\end{equation}
where
\begin{align}{\label{postbetaG1}}
s^2_j &= \frac{\sigma^2_e}{\bfx^T_j \bfx_j + \frac{\sigma^2_e}{\sigma^2_{\beta}}},  \nonumber \\
\mu_j &= \frac{\mathrm{\bf{x}}^T_j\mathbf{y} - \sum_{i \neq j}\mathbb{E}_q[\gamma_i \tilde{\beta}_i]\mathrm{\bf{x}}^T_j \mathrm{\bf{x}}_i}{\mathrm{\bf{x}}^T_j \mathrm{\bf{x}}_j + \frac{\sigma^2_e}{\sigma^2_{\beta}}},
\end{align}
and
\begin{align}\label{pi_j}
\pi_j = &\frac{1}{1 + \exp(-w_j)}, \nonumber \\
  w_j = &\log \frac{\pi}{1-\pi} + \frac{1}{2} \log \frac{s^2_j}{\sigma^2_{\beta}} + \frac{\mu_j^2}{2s^2_j} \nonumber \\
        &+ \sum_{k=1}^K \log \frac{\alpha_k p_{jk}^{\alpha_k -1 } u_k + 1 - u_k}{\alpha_k p_{jk}^{\alpha_k -1 } (1 -v_k) + v_k}.
\end{align}
Since $q(\tilde\bfbeta, \bfgamma, \bfGamma)$ is an approximation to the true posterior, the above result (\ref{q_j}) can be interpreted as follows. Here $\pi_j$ can be viewed as an approximation of $\pr(\gamma_j = 1|\bfy, \bfX, \bfP;\bftheta)$. \textcolor{black}{As we can observe from Eq.~\eqref{pi_j}, $\pi_j$ is influenced by three sources of information: the first term is the prior information of expected proportion of variants contributed to the variance of phenotype T, the second term $\frac{1}{2} \log \frac{s^2_j}{\sigma^2_{\beta}} + \frac{\mu_j^2}{2s^2_j}$ corresponds to the information provided by the genotype data and the third term $\displaystyle \sum_{k=1}^K \log \frac{\alpha_k p_{jk}^{\alpha_k -1 } u_k + 1 - u_k}{\alpha_k p_{jk}^{\alpha_k -1 } (1 -v_k) + v_k}$ corresponds to the information provided by the summary-level data. It is clear that the same stringent $p$-value in the GWAS with stronger pleiotropic relationship indicates a greater probabilistic association with the individual-level data.} On the other hand, when SNP $j$ is irrelevant to the phenotype ($\gamma_j = 0$), the approximated posterior of $\tilde{\beta}_j$ remains the same as its prior, i.e., $\tilde\beta_j \sim N(\tilde{\beta}_j|0,\sigma^2_\beta)$; When SNP $j$ is relevant ($\gamma_j = 1$), its posterior changes accordingly as $\tilde\beta_j \sim N(\tilde{\beta}_j|\mu_j,s^2_j)$.

With $q_j(\tilde{\beta}_j,\gamma_j,\bfGamma_j)$ given in (\ref{q_j}), the lower bound 
\begin{equation*}
\mathcal{L}(q) = \mathbb{E}_{q}[\log \pr(\bfy,\tilde{\bfbeta},\bfgamma,\bfGamma, \bfP |\bfX,\bftheta)] -\mathbb{E}_{q}[\log {q}(\tilde{\bfbeta},\bfgamma,\bfGamma)]
\end{equation*}
can be evaluated in a closed form. By taking derivative of $\mathcal{L}(q)$ with respect to each parameter in $\bftheta$ and setting them to zero, we can obtain the updating equations for parameter estimation (details in Section 1.2 of the Supplementary document).

In summary, our algorithm can be viewed as a variational expectation-maximization (EM) algorithm. In the expectation step, we evaluate the expectation with respect to the distribution $q$ to obtain the lower bound $\mathcal{L}(q)$ given in Eq.~(\ref{lowerbound}). In the maximization step, we maximize the current $\mathcal{L}(q)$ with respect to model parameters in $\bftheta$. Hence, the convergence of the proposed algorithm is guaranteed as the lower bound increases in each EM iteration.

\subsection{Identification of risk variants and risk prediction}
Given a genotype vector $\tilde\bfx = [\tilde{x}_1,\dots,\tilde{x}_M]^T$ of an individual, the well-trained LEP model conducts the risk prediction by computing
\begin{equation}\label{eq_pred}
  \hat{y} = c_0 + \sum_j \tilde{x}_j \mathbb{E}(\gamma_j \tilde{\beta}_j) = c_0 + \sum_j \tilde{x}_j \pi_j \mu_j.
\end{equation}
Here $\pi_j$ in Eq.~\eqref{eq_pred} is also an approximation to the true posterior $\pr(\gamma_j = 1| \bfy, \bfP,\bfX;\bftheta)$ for SNP $j$, as $\pr(\gamma_j = 0| \bfy, \bfP,\bfX;\bftheta)$ is the definition of local false discovery rate (FDR) of SNP $j$ (\cite{efron2010large}), we denote $fdr_j = 1 - \pi_j$ as the approximation of local FDR, SNP $j$ would be labeled as a risk variant if $fdr_j$ approaches $0$, e.g. $fdr_j \leq 0.05$.

\section{RESULTS}
\subsection{Simulation}
\begin{figure*}
  \centering
  \includegraphics[width=0.48\textwidth]{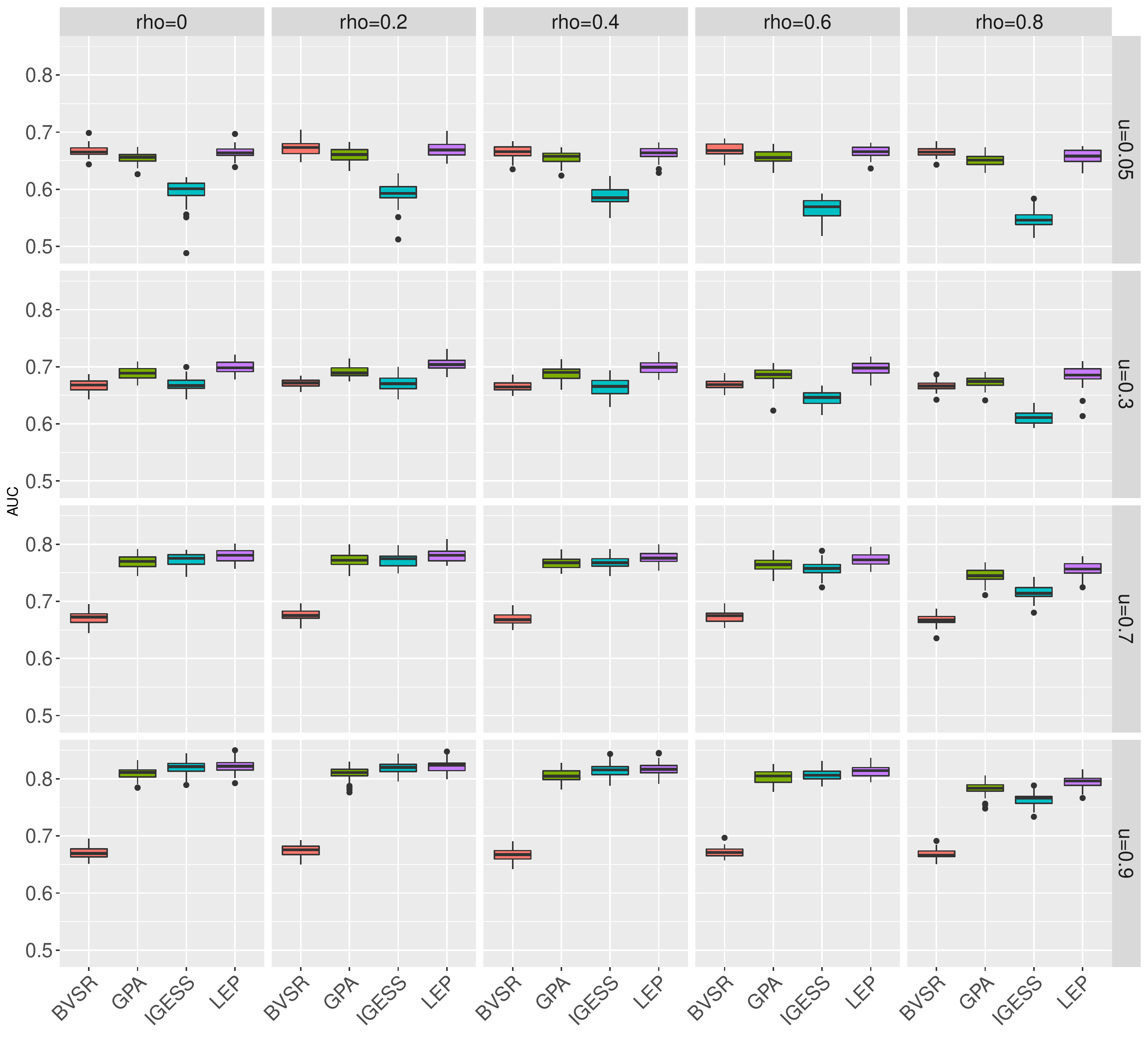} \quad
   \includegraphics[width=0.48\textwidth]{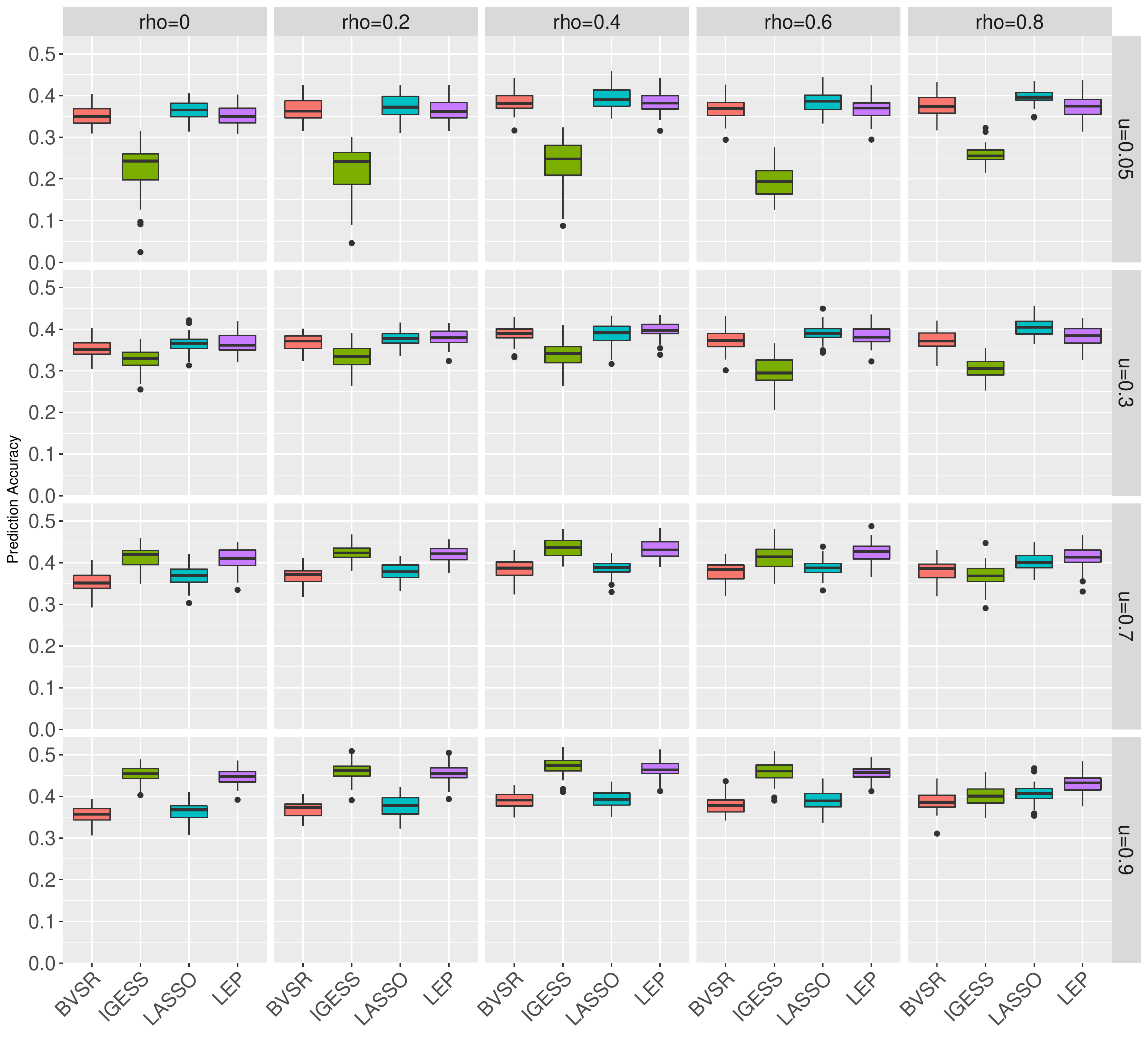} \\
  \caption{Comparison of BVSR, GPA, IGESS and Lasso with different $\{u, v\}$. Left panel: performance of risk variant identification measured by AUC; Right panel: performance of risk prediction. The number of summary-level datasets $K=1$. All the results are summarized based on 50 replications.}
  \label{fig:Along_u_rho}
 \end{figure*}

 \begin{figure*}
  \centering
    \includegraphics[width=0.48\textwidth]{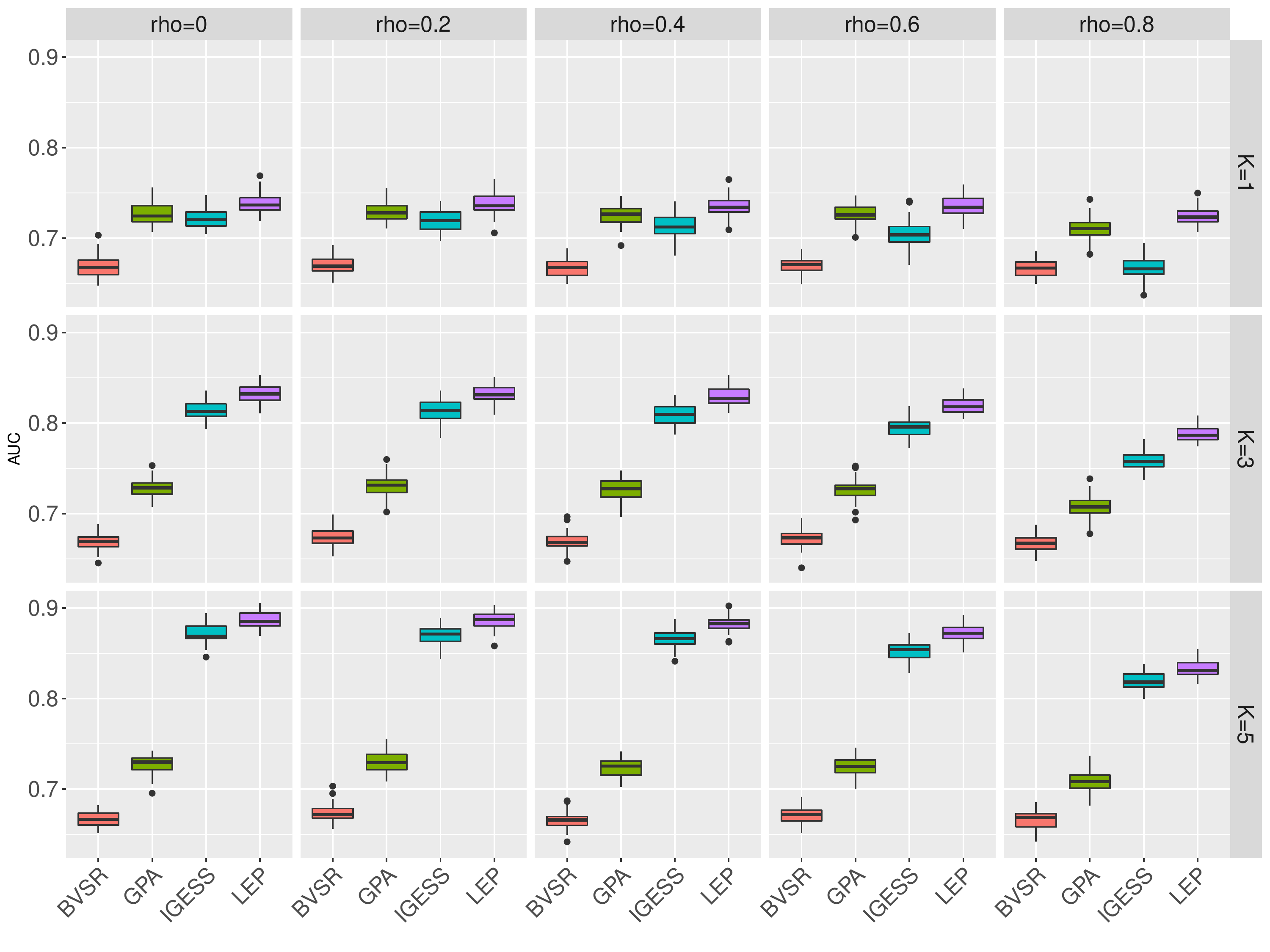}\quad
    \includegraphics[width=0.48\textwidth]{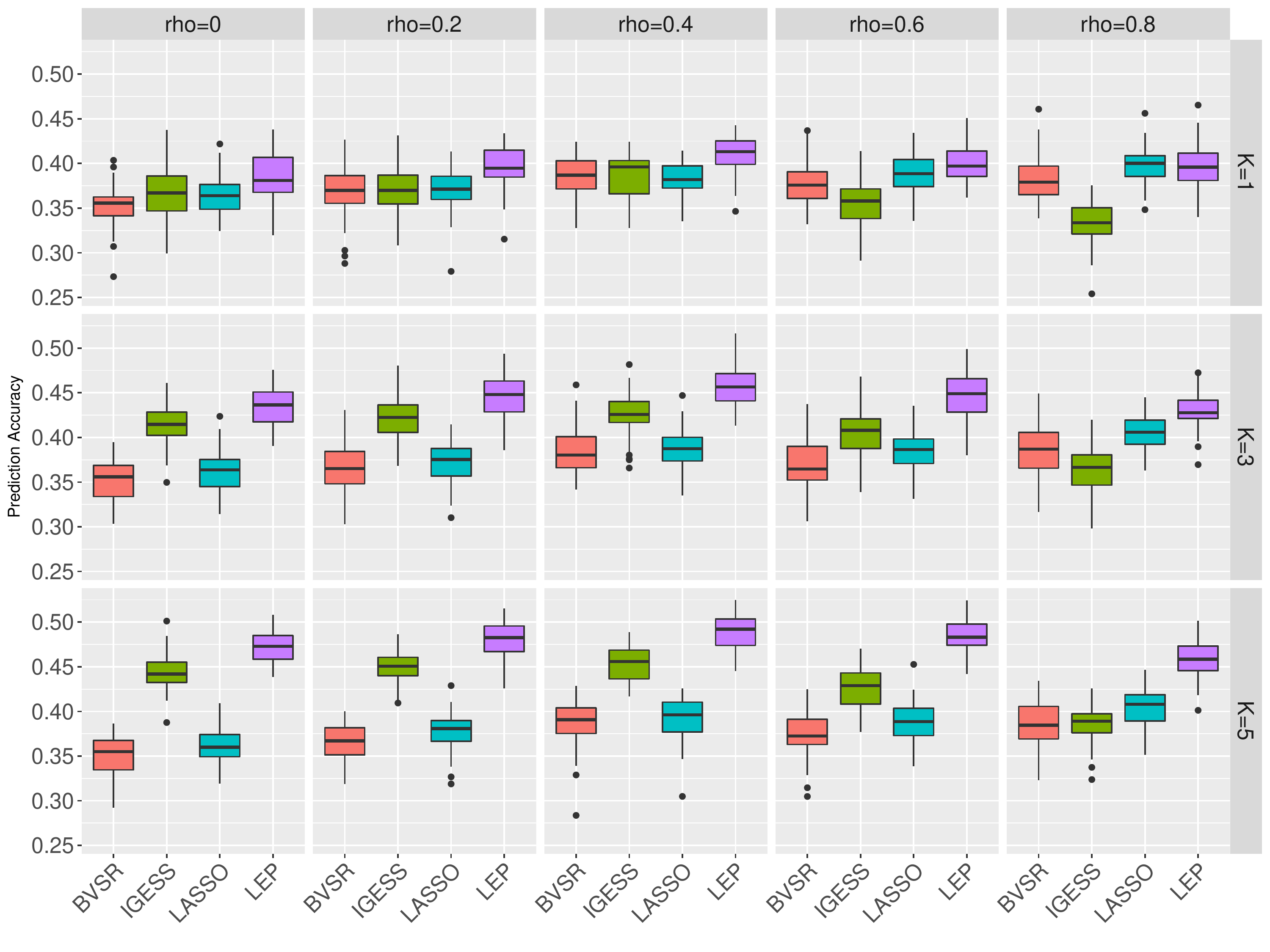}\\
  \caption{Comparison of BVSR, GPA, IGESS and Lasso with different $K$. Left panel: performance of risk variant identification measured by AUC; Right panel: performance of risk prediction. The indicator $u_k=0.5$ and sparsity-level $s=0.05$ . All the results are summarized based on 50 replications.}
  \label{fig:Along_rho_k}
 \end{figure*}

In this section, we conducted simulation studies to evaluate the performance of LEP in terms of both the risk variants identification and the risk prediction. The performance of the identification of risk variants was evaluated in comparison with BVSR \citep{carbonetto2012scalable}, Lasso \citep{tibshirani1996regression} (only with individual-level data), IGESS \citep{dai2017igess} and GPA \citep{chung2014gpa}. Note that IGESS was based on a homogeneous assumption, i.e. $\Gamma_j^1$ = $\Gamma_j^2$ = \ldots = $\Gamma_{jk}$ = $\gamma_j$, while GPA only worked for summary-statistics data. Then, the prediction accuracy of LEP was evaluated in comparison with BVSR, Lasso and IGESS. In the spirit of reproducibility, all the
simulation codes were made publicly available at \url{https://github.com/daviddaigithub/LEP}.



\subsubsection{Simulation settings}
Both the individual-level and summary-level data were simulated in our experiments. In order to mimic the linkage disequilibrium (LD) of the individual-level data, the genotype matrix $\bfX$ was first generated by a zero-mean Gaussian distribution with a autoregressive structure for the co-variance matrix, where the autoregressive correlation $\rho^{|j-j'|}$ indicated the correlation between the two corresponding variants. Clearly, $\rho = 0$ corresponded to the situation that all the variants were independent. We considered $\rho = \{0,0.2,0.4,0.6,0.8\}$ in our simulations. Then each column of $\bfX$ was numerically coded as $\{0, 1, 2\}$ with probability $(1-f)^2, 2f(1-f)$ and $f^2$, respectively, according to Hardy-Weinberg principle, where $f$ was the minor allele frequency (MAF). We fixed the number of samples as $N=2,000$, the number of variants as $M=10,000$, and the sparsity-level $s = 0.05$. On average, $M\cdot s$ nonzero entries in the vector of effect sizes $\bfbeta$ were generated using $N(0,1)$. The phenotype vector was generated as $\bfy = \bfX \bfbeta + \bfe$, where the variance of $\bfe$ was adjusted such that heritability, i.e., $\mathrm{var}(\bfX\bfbeta)/\mathrm{var}(\bfy)$ was fixed at 50\%.




The simulated $p$-values were obtained from the individual-level data instead of the generative model (Eq.~\eqref{eqpriorpvalue}). \textcolor{black}{Because the released summary-level data  sets often have much larger sample sizes, we used $n_0= 8,000$ samples to generate summary statistics, i.e., $p$-values in our simulation.}  
We generated individual-level genotype data $\bfX^{(k)}$, as what we described above. To simulate the vector of effect size $\bfbeta^{(k)}$, we assume the $k$-th trait has the same of proportion of nonzero entries as the trait T, that is $s^k = s$. We generated the nonzero entries according to $(u_k, v_k)$ in Eq.~\eqref{uv} where $(u_k, v_k)$ satisfied $u_k s + (1-v_k) (1-s) = s$, i.e., probability relationship,
\begin{equation*}
\begin{split}
 \pr(\Gamma_{jk} = 1)=&\pr(\Gamma_{jk}=1|\gamma_j=1)Pr(\gamma_j=1) \\
                  &+ \pr(\Gamma_{jk}=1|\gamma_j=0)Pr(\gamma_j=0).
\end{split}
 \end{equation*} 
Again, the nonzero entries were drawn from $N(0,1)$, and the $k$-th phenotype vector was generated as $\bfy^{k} = \bfX^{k} \bfbeta^{k} + \bfe_k$, where the variance of $\bfe_k$ was specified such that the heritability of the $k$-th phenotype was $50\%$. Then we obtained $p$-values of each variant by applying simple linear regression. To evaluate the performance of different methods, we pretended that we can not access the individual level data set but can only use the summary-level data, i.e. $p$-value matrix $\bfP\in \mathbb{R}^{M\times K}$.


 For LEP and IGESS, both the individual-level data $\{\bfX,\bfy\}$ and the $p$-value matrix $\bfP$ were used. For BVSR and Lasso, only the individual-level data set $\{\bfX,\bfy\}$ was used, and their performance could serve as a baseline. For GPA, an $M\times 2$ matrix $\bfP$ containing the $p$-values from $\{\bfX,\bfy\}$ and the $p$-values from the first study $\{\bfX^{(1)}, \bfy^{(1)}\}$ was used as its input. We present the results of analyzing quantitative phenotype simulated as above in the main text and leave the results of case-control study in the Supplementary document.




\subsubsection{Results}
We conducted the comparisons in four scenarios and evaluated the performance of both risk variants identification and risk prediction. The performance of risk variants identification was measured by the area under the receiver operating characteristic (ROC) curve (AUC), and the performance of risk prediction was measured by the correlation between the observed phenotype values and the predicted values. 

First, we fixed $K=1$, and varied $u$ at $\{0.9,0.7,0.3,0.05\}$ to mimic different pleiotropic settings to see the benefits of LEP. According to the condition of independence ($u$+$v$=$1$, see the detailed derivation in the Supplementary document), the parameter setting $u$ = 0.05 ($v$ = 0.95) corresponds to the situation that the trait of the summary-level data are irrelevant to the trait of the individual-level data. The pleiotropic effect between two traits increases as $u$ increases from $0.3$ to $0.9$. Fig.~\ref{fig:Along_u_rho} shows the performance of risk variants identification and prediction under different pleiotropy settings. Here BVSR and Lasso using only individual-level data were used to serve as a baseline. In terms of risk variants identification, when the pleiotropic effect between two traits is high ($u$ = $\{0.7,0.9\}$), LEP, GPA and IGESS outperform BVSR working only with the individual-level dataset; when the pleiotropic effect is weak ($u=0.3$), LEP and GPA still perform better than BVSR. Moreover, when there is no relevant summary information ($u=0.05$), the performance of LEP is comparable to BVSR, which shows the robustness of LEP in the absence of useful summary information. It has similar results in terms of risk prediction. LEP and IGESS outperform BVSR and Lasso when the pleiotropic effect is high ($u$ = $\{0.7,0.9\}$). When the pleiotropic effect is weak ($u$ = $0.3$), LEP outperforms BVSR; When the summary-level data is irrelevant, LEP is slightly worse than Lasso, but is comparable with BVSR.

Second, we fixed 
$u_k := \pr(\Gamma_{jk} = 1| \gamma_j = 1) = 0.5$, and varied the number of study with summary statistics at $K=\{1,3,5\}$ to see the improvements when more information was incorporated. Fig.~\ref{fig:Along_rho_k} shows the improvement of LEP in comparisons with GPA, IGESS, BVSR and Lasso as $K$ increases. When $K = 1$, GPA, IGESS and LEP all outperform BVSR as they use more information. LEP outperforms GPA as there is more useful information from the individual-level data. The performance of LEP increases as the number of $K$ increases. One can observe that IGESS has comparable performance to LEP but it can not control the FDR as shown in the Supplementary document. Instead, LEP is a method that could borrow strength from the study of a pleiotropic relationship and also control the FDR at the pre-specified level. All the results with respect to FDR, the power of risk variants identification are presented in the Supplementary document.

Last, we considered an extreme situation, the association status $\bfGamma_{\cdot k}$ (the $k$-th column of $\bfGamma$) was exactly the same as 
$\bfgamma$, which was the assumption of IGESS, and the results are presented in the Supplementary document.

\begin{figure}[h!]
    \includegraphics[width=0.49\textwidth]{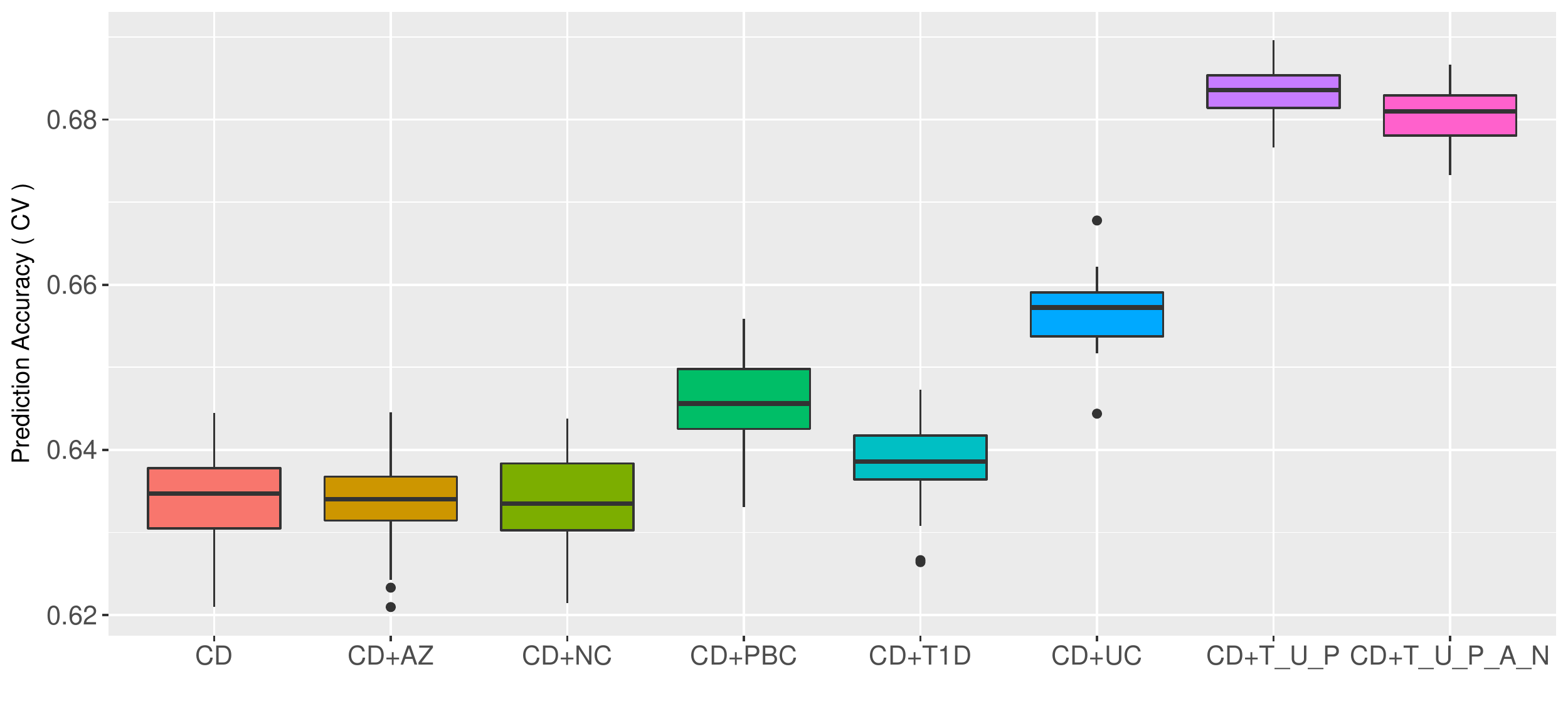}
    \centering
    \caption{Prediction accuracy measured by AUC with different summary-level data. The plus sign `+' means jointly analyzing CD and summary-level data for certain trait (s); T$\_$U$\_$P means jointly analyzing with T1D, UC and PBC; T$\_$U$\_$P$\_$A$\_$N means jointly analyzing with T1D, UC, PBC, AZ and NC. The results are based on 30 replications.}
    \label{fig:LEP_REAL}
\end{figure}

\begin{figure*}
  \centering
  \includegraphics[width=0.95\textwidth]{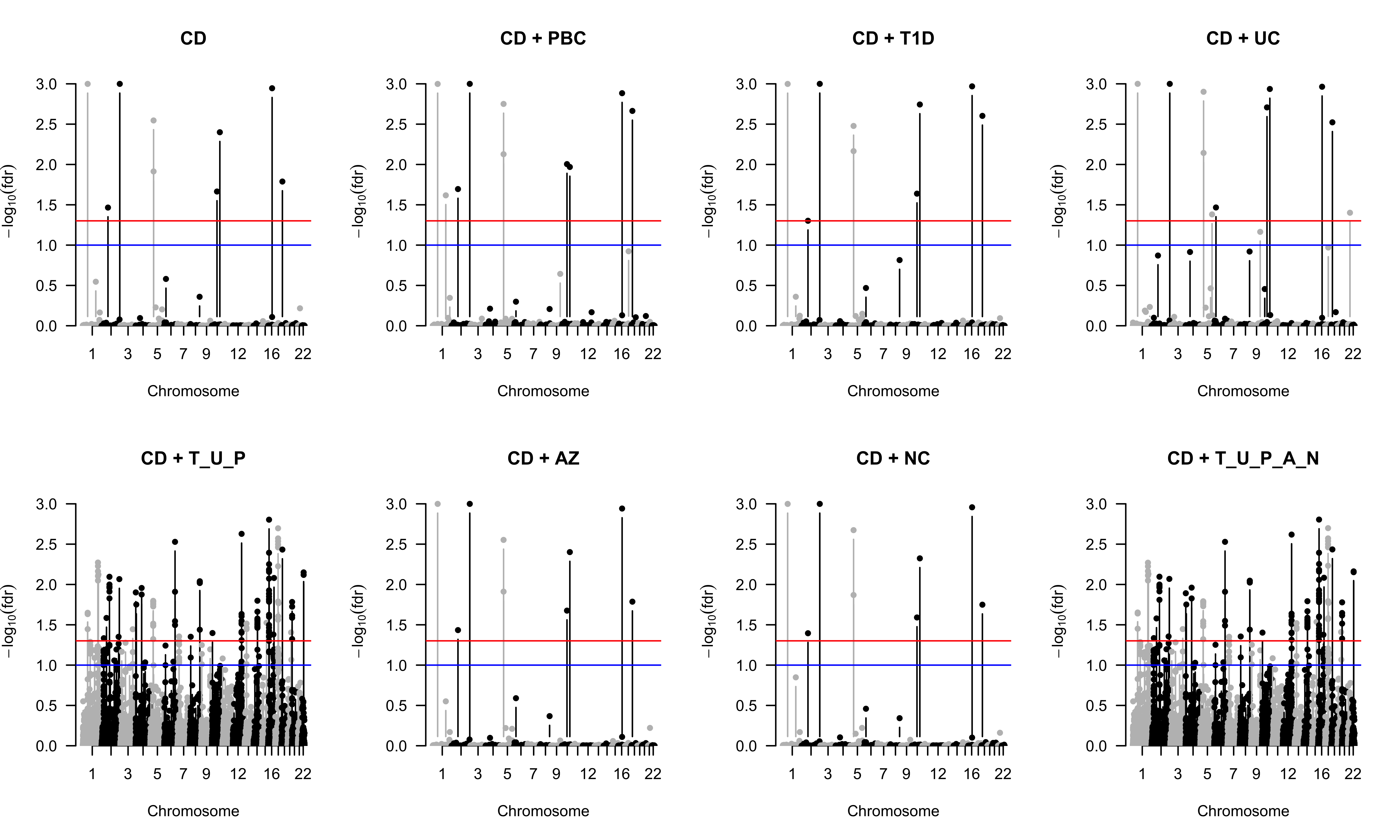}\\
  \caption{Manhattan plots of the analysis results of Crohn's Disease using LEP. T$\_$U$\_$P means jointly analyzing with T1D, UC and PBC; T$\_$U$\_$P$\_$A$\_$N means jointly analyzing with T1D, UC, PBC, AZ and NC. The red lines and blue lines correspond to $fdr=0.05$ and $fdr=0.1$ respectively.}
  \label{fig:man2}
\end{figure*}

In summary, all the simulation results suggest that LEP is a reliable method when incorporating information from various kinds of summary statistics. It could make the most efficient use of the information as well as control the FDR well. Moreover, it will not degenerate the performance of analyzing individual-level data even when the summary statistics are from the irrelevant studies.

\subsection{Real Data Analysis}
We applied LEP to analyze GWAS data of Crohn's disease (CD) with summary statistics from some other autoimmune diseases. The individual-level data is from the Welcome Trust Case Control Consortium (WTCCC) \citep{burton2007genome}. There are 5,009 samples, of which 2,005 are cases and 3,004 are controls. The summary-level datasets are for Ulcerative colitis (UC), Primary biliary cirrhosis (PBC), Type 1 diabetes (T1D), three autoimmune-type diseases, are expected to be helpful in jointly analyzing with CD. In order to demonstrate the robustness of LEP on real data analysis, we selected two GWASs for Alzheimer (AZ), Neuroticism (NC), which were not relevant to CD by our common sense.

\begin{table}[tp]
\caption{Pleiotropy for the five data set with {\emph{u},\emph{v}}} \label{tab:a}
\centering
\begin{footnotesize}
\begin{tabular}{rlcc}
  \hline
 & GWAS & \emph{u} & \emph{v} \\ 
  \hline
  1 & Ulcerative Colitis & 1.0000 & 0.9260 \\ 
  2 & Primary biliary cirrhosis & 0.9998 & 0.9744 \\ 
  3 & Type 1 diabetes & 0.9986 & 0.9176 \\ 
  4 & Alzheimer & 2.72e-05 & 0.9325 \\ 
  5 & Neuroticism & 2.32e-03 & 0.7759 \\
   \hline
\end{tabular}
\end{footnotesize}
\end{table}

\subsubsection{Quality Control on the individual-level data}
The strict quality control (QC) on the individual-level data from WTCCC was performed by using PLINK \citep{purcell2007plink} and GCTA \citep{yang2011gcta}. First, we removed the individuals that has more than 2\% missing genotypes; Before combining the CD dataset and the two control datasets, the SNPs with MAFs $<$ 0.05 and those with missing rate $>1\%$ were removed. Then we removed the SNPs with $p$-value $<0.0001$ in Hardy-Weinberg equilibrium test and kept one of the relatives with estimated relatedness $>0.025$ identified by GCTA \citep{yang2011gcta} on the combined dataset. After QC, there were 1,656 cases and 2,880 controls remaining with 308,950 SNPs.

\subsubsection{Summary-statistic data preprocessing}
The details of the summary-level data were listed in Table S1 in Supplementary document. Taking overlapped SNPs among individual-level data and all the five phenotypes led to the individual-level data $\bfX \in \mathbb{R}^{N \times M}, 
\bfy \in \mathbb{R}^{N \times 1}$ and a $p$-value matrix $\bfP \in \mathbb{R}^{M \times K}$ , where $N = 4,536$, $M=194,782$ and $K=5$.

\subsubsection{Analysis of pleiotropy with each single phenotype}
We applied LEP to analyze the individual-level data for Crohn's Disease with each of the five phenotypes. The results are listed in Table \ref{tab:a}. The greater value of $u$ indicates a stronger pleiotropic relationship between the corresponding disease and the Crohn's Disease. The values $\{u, v\}$ for Alzheimer and Neuroticism indicate that they almost share no associated variants with Crohn's disease.

\subsubsection{Analysis result of Crohn's Disease}
After analyzing the pleiotropy of each dataset with summary statistics, we then applied LEP to analyze CD with the selected summary statistics of PBC, T1D, UC, AZ and NC. The results for prediction accuracy and Manhattan plots of genome-wide hits are shown in Fig.~\ref{fig:LEP_REAL} and Fig.~\ref{fig:man2}, respectively. 

The Manhattan plots in Fig.~\ref{fig:man2} indicate that it is more effective in identifying risk variants for CD by integrating the summary statistics from auto-immune diseases, i.e. UC, PBC and T1D, especially for combining them all. The irrelevant traits AZ, NC did not degenerate the performance of LEP in terms of identifying risk causal variants.  

The results in Fig.~\ref{fig:LEP_REAL} imply that the summary statistics of PBC, T1D and UC offer different degrees of relevant information, improving prediction accuracy (measured by AUC) from 63.39\% ($\pm$0.58\%) (only CD) to 65.66\% ($\pm$0.44\%) (with UC), and combining them all together could improve the prediction accuracy to 68.33\% ($\pm$0.32\%). These results suggested that information from diseases of a pleiotropic relationship could indeed substantially contribute to genetic prediction accuracy.  

To demonstrate the robustness of LEP on the real data, we applied LEP on the individual-level data with each of them, their performance (63.34\% ($\pm$0.53\%), 63.40\% ($\pm$0.54\%)) was comparable with that of working on individual-level data alone (63.39\% ($\pm$0.58\%)). We also applied LEP with all the original summary statistics PBC, T1D, UC, AZ and NC, the performance slightly degraded from 68.33\% ($\pm$0.32\%) to 68.05\% ($\pm$0.34\%) suggesting the robustness of LEP when irrelevant summary-level datasets were included.


\section{Conclusion}
Restricted access to the individual-level data and the advantages of using summary-level data motivate the recent development of many new methods for further analyzing summary-level data. Working with the available but limited individual-level data and the abundant summary-level data is a promising direction to make the most efficient use of existing data resources. In this study, we propose a statistical approach, LEP, to integrating data at both the individual-level and summary-level, and the proposed method build a bridge from the individual-level data to summary-level data by modeling the pleiotropy. The implemented variational inference makes LEP scalable to genome-wide data analysis. The comprehensive simulations and real data analysis demonstrate the advantages of LEP and its wide applicabilities in practice. LEP will serve as a powerful tool for jointly analyzing individual-level data and summary statistics.

\section{Appendix}
\subsection{Variational Expectation-Maximization Algorithm}
\subsubsection{E-Step}
  The joint probability for the LEP model could be rewritten as ,
      let $\theta = \{\sigma^2_{\beta},\sigma^2_{e},\pi, \{\alpha_k\}_{k=1}^K\}$ be the collection of model parameters.
      \begin{equation}{\label{eqjoint2}}
      \begin{split}
       &\pr(\bfy,\tilde{\bfbeta},\bfgamma, \bfGamma,\bfP |\bfX;\bftheta)  \\
       = &\pr(\bfy|\tilde{\bfbeta},\bfgamma, \bfX;\bftheta) \pr(\tilde{\bfbeta};\bftheta) \pr(\bfgamma|\bftheta)\pr(\mathbf{P}|\bfGamma;\bftheta) \pr(\bfGamma | \bfgamma).  \\
      \end{split}
      \end{equation}

We model the relationship between $\gamma_j$ and $\Gamma_{jk}$ as

\begin{equation}
  \begin{split}
    u_k := \pr(\Gamma_{jk} = 1| \gamma_j = 1), \\
        v_k := \pr(\Gamma_{jk} = 0| \gamma_j = 0).  \\
  \end{split}
\end{equation}
For the clarity of expression, we divide the joint probability into two parts, $\pr(\bfy|\tilde{\bfbeta},\bfgamma, \bfX;\bftheta) \pr(\tilde{\bfbeta};\bftheta) \pr(\bfgamma|\bftheta)$, for the individual-level data and $\pr(\mathbf{P}|\bfGamma;\bftheta) \pr(\bfGamma | \bfgamma)$ for the summary-level data. We expanded them in the following form,
\begin{equation}
   \pr(\bfy|\tilde{\bfbeta},\bfgamma, \bfX;\bftheta) \pr(\tilde{\bfbeta};\bftheta) \pr(\bfgamma|\bftheta) =  N(\bfy|\sum_j  \bfx_j \tilde{\beta}_j  \gamma_j,\sigma^2_e \mathrm{I}) \prod_{j=1}^{M} N(\tilde{\beta}_j | 0,\sigma^2_{\beta})  \pi ^{\gamma_j} (1-\pi) ^{1-\gamma_j}
\end{equation} and
\begin{equation}
    \begin{split}
    \pr(\mathbf{P}|\bfGamma;\bftheta) \pr(\bfGamma | \bfgamma) &= \prod_{j=1}^M   \left (
    \prod_{k=1}^K \left(\pr(p_{jk}|\Gamma_{jk} = 1)\pr(\Gamma_{jk} = 1|\gamma_j = 1)\right)^{\Gamma_{jk}}  \left( \pr(p_{jk}|\Gamma_{jk} = 0)\pr(\Gamma_{jk} = 0|\gamma_j = 1)\right)^{1 - \Gamma_{jk}}  \right ) ^{\gamma_j }   \\
    &   \left (
    \prod_{k=1}^K \left(\pr(p_{jk}|\Gamma_{jk} = 1)\pr(\Gamma_{jk} = 1|\gamma_j = 0)\right)^{\Gamma_{jk}}  \left( \pr(p_{jk}|\Gamma_{jk} = 0)\pr(\Gamma_{jk} = 0|\gamma_j = 0)\right )^{1 - \Gamma_{jk}}  \right ) ^{1 - {\gamma_j} } \\
    &= \prod_{j=1}^M  \left (
    \prod_{k=1}^K (\alpha_k p_{jk}^{\alpha_k -1} u_k)^{\Gamma_{jk}}  (1-u_k)^{1 - \Gamma_{jk}}  \right ) ^{\gamma_j }  \left (
    \prod_{k=1}^K (\alpha_k p_{jk}^{\alpha_k -1} (1-v_k))^{\Gamma_{jk}}  (v_k)^{1 - \Gamma_{jk}}  \right ) ^{1-\gamma_j}. \\
    \end{split}
\end{equation}

The logarithm of the marginal likelihood is
\begin{equation}{\label{eqlower}}
  \begin{split}
  \log \pr(\bfy,\mathbf{P}|\bfX;\bftheta) &= \log \sum_{\bfgamma} \sum_{\bfGamma} \int_{\tilde{\bfbeta}} \pr(\bfy,\mathbf{P},\tilde{\bfbeta},\bfgamma, \bfGamma|\bfX,\bftheta)  d\tilde{\bfbeta}  \\
  & \geq \sum_{\bfgamma}  \sum_{\bfGamma} \int_{\tilde{\bfbeta}} q(\tilde{\bfbeta},\bfgamma,\bfGamma) \log \frac{\pr(\bfy,\tilde{\bfbeta}, \mathbf{P},\bfgamma, \bfGamma |\bfX,\bftheta) }{ q(\tilde{\bfbeta},\bfgamma,\bfGamma)} d\tilde{\bfbeta}  \\
  & = \mathbb{E}_q[\log \pr(\bfy,\mathbf{P},\tilde{\bfbeta},\bfgamma,\bfGamma|\bfX;\bftheta)  - \log q(\tilde{\bfbeta},\bfgamma,\bfGamma) ] \\
  &:= \mathcal{L}(q),
  \end{split}
\end{equation}
where $\mathcal{L}(q)$ is the lower bound implied by Jensen's inequality and the equality holds if and only if $q(\tilde{\bfbeta},\bfgamma,\bfGamma)$ is the true posterior $\pr(\tilde{\bfbeta},\bfgamma,\bfGamma|\bfy,\bfP,\bfX;\bftheta)$. Instead of working with the marginal likelihood, we iteratively maximizing $\mathcal{L}(q)$. As it is stated in the main text, we employ the following variational distribution to make it feasible to evaluate the lower bound,
\begin{equation}
  q(\tilde{\bfbeta},\bfgamma,\bfGamma) = \prod_{j=1}^M q_j(\tilde{\bfbeta}_j,\gamma_j,\bfGamma_j),
\end{equation}
where $\bfGamma_j = \{\Gamma_{j1},\hdots,\Gamma_{jK}\}$. According to the nice property of factorized distributions in variational inference, we can obtain the best approximation as 
\begin{equation}{\label{eqex}}
  \log q_j(\tilde{\beta}_j , \gamma_j, \bfGamma_j) = \mathbb{E}_{i\neq j}[\log \pr(\bfy,\bfP,\tilde{\bfbeta},\bfgamma,\bfGamma|\bfX;\bftheta)] + \mathrm{Const},
\end{equation}
where the expectation is taken with respect to all of the other factors $\{q_i(\tilde{\beta}_i,\gamma_i, \bfGamma_i)\}$ for $i \neq j$.

The logarithm of the joint probability function is
\begin{equation}{\label{eqjoint}}
  \begin{split}
    \log \pr(\bfy,\bfP,\tilde{\bfbeta},\bfgamma, \bfGamma|\bfX;\bftheta) &= -\frac{N}{2} \log(2\pi \sigma^2_e) - \frac{\bfy^T\bfy}{2 \sigma^2_e} \\
     &\quad + \frac{\sum_{j=1}^{M}\gamma_j \tilde{\beta}_j \bfx^T_j \bfy}{\sigma^2_e} -\frac{1}{2\sigma^2_e} \sum_{j=1}^{M} \left( (\gamma_j \tilde{\beta}_j)^2 \bfx_j^T \bfx_j\right)\\
     &\quad -\frac{1}{2\sigma^2_e} \left( \sum_{j=1}^M \sum_{j' \neq j}^{M} \gamma_j \tilde{\beta}_j \gamma_{j'} \tilde{\beta}_{j'} \bfx_k \bfx_k\right) \\
     & \quad -\frac{M}{2} \log(2\pi \sigma^2_{\beta}) - \frac{1}{2\sigma^2_{\beta}} \sum_{j=1}^{M} \tilde{\beta}_j^2 \\
     &\quad + \log \pi \sum_j \gamma_j  + \log(1-\pi) \sum_j(1-\gamma_j)\\
     &+ \sum_{j=1}^M \gamma_j \sum_{k=1}^K \Gamma_{jk} \log(\alpha_k p_{jk}^{\alpha_k-1} u_k) + (1-\Gamma_{jk})\log(1-u_k)  \\
     & + \sum_{j=1}^M (1-\gamma_j) \sum_{k=1}^K \Gamma_{jk}\log(\alpha_k p_{jk}^{\alpha_k-1} (1-v_k)) + (1-\Gamma_{jk})\log(v_k).
  \end{split}
\end{equation}

Before proceeding, we should keep several things in our mind. First, $q(\tilde{\bfbeta},\bfgamma,\bfGamma)$ is the variational approximation to the posterior $\pr(\tilde{\bfbeta},\bfgamma,\bfGamma |\bfy,\bfX;\bftheta)$. Second, we assumed $q(\tilde{\bfbeta},\bfgamma,\bfGamma) =\displaystyle \prod_{j=1}^{M} q(\tilde{\beta}_j,\gamma_j,\bfGamma_j)$. Third, $q(\tilde{\beta}_j,\gamma_j,\bfGamma_j)=q(\tilde{\beta}_j|\gamma_j) q(\bfGamma_j|\gamma_j) q(\gamma_j)$.

To take the expectation in \eqref{eqex}, we rearrange \eqref{eqjoint} into the terms with and without index $j$
\begin{equation}{\label{eqsplit}}
\begin{split}
 \log \pr(\bfy,\bfP,\tilde{\bfbeta},\bfgamma,\bfGamma|\bfX;\bftheta)  &= -\frac{N}{2} \log(2\pi \sigma^2_e) - \frac{\bfy^T\bfy}{2 \sigma^2_e} \\
&\quad +\frac{\gamma_j \tilde{\beta}_j \bfx^T_j \bfy}{\sigma^2_e} + \frac{\sum_{k \neq j}\gamma_k \tilde{\beta}_k \bfx^T_j \bfy}{\sigma^2_e} \\
&\quad -\frac{1}{2\sigma^2_e} \left( (\gamma_j \tilde{\beta}_j)^2 \bfx^T_j \bfx_j\right) -\frac{1}{2\sigma^2_e} \sum_{k \neq j} \left( (\gamma_k \tilde{\beta}_k)^2 \bfx^T_k \bfx_k\right)\\
&\quad -\frac{1}{\sigma^2_e} \left( \sum_{k\neq j}^{M} \gamma_j \tilde{\beta}_j \gamma_{k} \tilde{\beta}_{k} \bfx^T_j \bfx_k\right) -\frac{1}{2\sigma^2_e} \left( \sum_{k\neq j} \sum_{k' \neq j} \gamma_k \tilde{\beta}_k \gamma_{k'} \tilde{\beta}_{k'} \bfx^T_k \bfx_{k'}\right) \\
& \quad -\frac{M}{2} \log(2\pi \sigma^2_{\beta})  - \frac{1}{2\sigma^2_{\beta}} \tilde{\beta}_j^2 - \frac{1}{2\sigma^2_{\beta}} \sum_{k\neq j} \tilde{\beta}_k^2 \\
 &\quad + \log \pi \sum_j \gamma_j  + \log(1-\pi) \sum_j(1-\gamma_j)\\
     &+ \sum_{j=1}^M \gamma_j \sum_{k=1}^K \Gamma_{jk}\log(\alpha_k p_{jk}^{\alpha_k-1} u_k) + (1-\Gamma_{jk})\log(1-u_k)  \\
     & + \sum_{j=1}^M (1-\gamma_j) \sum_{k=1}^K \Gamma_{jk} \log(\alpha_k p_{jk}^{\alpha_k-1} (1-v_k)) + (1-\Gamma_{jk})\log(v_k).
\end{split}
\end{equation}

Because
\begin{equation}
\pr(\Gamma_{jk} | p_{jk},\gamma_j) = \frac{\pr(p_{jk}|\Gamma_{jk})\pr(\Gamma_{jk}|\gamma_j)}{\pr(p_{jk}|\Gamma_{jk}=1)\pr(\Gamma_{jk} = 1|\gamma_j) + \pr(p_{jk}|\Gamma_{jk} = 0)\pr(\Gamma_{jk} = 0|\gamma_j)},  
\end{equation}  
we have
\begin{equation}
\begin{split}
\pr(\Gamma_{jk} = 1| p_{jk},\gamma_j = 1) &= \frac{\alpha_k p^{\alpha_k - 1}_{jk} u_k}{\alpha_k p^{\alpha_k - 1}_{jk} u_k + (1-u_k)}  =: q^{jk}_{11},  \\
\pr(\Gamma_{jk} = 0| p_{jk},\gamma_j = 1) &= \frac{1 - u_k}{\alpha_k p^{\alpha_k - 1}_{jk} u_k + (1-u_k)} =: q^{jk}_{10},  \\
\pr(\Gamma_{jk} = 1| p_{jk},\gamma_j = 0) &= \frac{\alpha_k p^{\alpha_k - 1}_{jk} (1-v_k)}{\alpha_k p^{\alpha_k - 1}_{jk} (1-v_k) + v_k}  =: q^{jk}_{01}, \\
\pr(\Gamma_{jk} = 0| p_{jk},\gamma_j = 0) &= \frac{v_k}{\alpha_k p^{\alpha_k - 1}_{jk} (1-v_k) + v_k}  =: q^{jk}_{00}.
\end{split}
\end{equation}

Now we can take expectation of $\log \pr(\bfy,\mathrm{P},\tilde{\bfbeta},\bfgamma,\bfGamma|\bfX;\bftheta)$ under the distribution $q(\tilde{\beta}_{-j},\bfGamma_{-j.},\gamma_{-j}).$
When $\gamma_j = 1$, we have
\begin{equation}
  \log q(\tilde{\beta}_j |\gamma_j = 1) = \left( -\frac{1}{2\sigma^2_e}\bfx^T_j \bfx_j - \frac{1}{2\sigma^2_{\beta}}\right) \tilde{\beta}^2_{j} + \frac{\bfx^T_j \bfy - \displaystyle \sum_{k\neq j}^{M} \mathbb{E}_k[\gamma_k \tilde{\beta}_k] \bfx^T_j \bfx_k}{\sigma^2_e}\tilde{\beta}_j + \mathrm{Const}
\end{equation}  
and
\begin{equation}
  q(\Gamma_{jk} | \gamma_j = 1) =   (q_{11} ^{jk})^{\Gamma_{jk}}  (q^{jk}_{10}) ^{1 - \Gamma_{jk}}.
\end{equation}

Because $\log q(\tilde{\beta}_j |\gamma_j = 1) $ is  a quadratic form, we know $q(\beta_j | \gamma_j = 1) = N(\beta_j | \mu_j,s^2_j)$, where
\begin{equation}{\label{postbetaG1}}
  \begin{split}
   s^2_j &= \frac{\sigma^2_e}{\bfx^T_j \bfx_j + \frac{\sigma^2_e}{\sigma^2_{\beta}}},  \\
   \mu_j &= \frac{\bfx^T_j \bfy - \sum_{k \neq j}\mathbb{E}_k[\gamma_k \beta_k]\bfx^T_j \bfx_k}{\bfx^T_j \bfx_j + \frac{\sigma^2_e}{\sigma^2_{\beta}}}.
  \end{split}
\end{equation}

Similarly, when $\gamma_j = 0$, we have
\begin{equation}
\log q(\tilde{\beta}_j |\gamma_j = 0) = -\frac{1}{2\sigma^2_{\beta}}\tilde{\beta}^2_j + \mathrm{Const} \label{postbetaG2}
\end{equation}
and
\begin{equation}
q(\Gamma_{jk} | \gamma_j = 0) = (q_{01} ^{jk})^{\Gamma_{jk}}  (q^{jk}_{00}) ^{1 - \Gamma_{jk}}.
\end{equation}

According to Eq.~\eqref{postbetaG2}, we know $q(\tilde{\beta}_j|\gamma_j = 0) = N(\tilde\beta_j|0,\sigma^2_{\beta})$. This is a very good property as it says that the posterior distribution of $\tilde{\beta}_j$ will be the same as its prior if this variable is irrelevant ( $\gamma_j = 0$ ). Note that $\gamma_j$ is a binary variable and then denote $\pi_j = q(\gamma_j = 1).$ Therefore we have

\begin{equation}
  q(\tilde{\beta}_j,\gamma_j,\bfGamma_j) = \left(\pi_j N(\mu_j, s^2_j) \prod_{k=1}^K q(\Gamma_{jk}|r_j = 1) \right)^{\gamma_j} \left((1-\pi_j)N(0,\sigma^2_{\beta})  \prod_{k=1}^K q(\Gamma_{jk}|r_j = 0)  \right)^{1-\gamma_j}.
\end{equation}

Now we evaluate the variational lower bound $\mathcal{L}(q)$  \eqref{eqlower}.

\begin{equation}\label{eq:lq1}
  \begin{split}
  \mathcal{L}(q) =  \mathbb{E}_q[\log \pr(\bfy,\mathbf{P},\tilde{\bfbeta},\bfgamma,\bfGamma|\bfX;\bftheta)]  - \mathbb{E}_q[ \log q(\tilde{\bfbeta},\bfgamma,\bfGamma) ]. \\
  \end{split}
\end{equation}

The first part of Eq.~\eqref{eq:lq1} could be written as
\begin{equation}
\begin{split}
&\quad \mathbb{E}_q[\log \pr(\bfy,\mathbf{P},\tilde{\bfbeta},\bfgamma,\bfGamma|\bfX;\theta)] \\
&= -\frac{N}{2} \log(2\pi \sigma^2_e) - \frac{\bfy^T\bfy}{2 \sigma^2_e} \\
&+ \frac{\displaystyle \sum_{j = 1}^M \mathbb{E}[\gamma_j \tilde{\beta}_j] \bfx^T_j \bfy}{\sigma^2_e} \\
&-\frac{1}{2\sigma^2_e} \sum_{j=1}^M \left( \mathbb{E}[(\gamma_j \tilde{\beta}_j)^2] \bfx^T_j \bfx_j\right)\\
&-\frac{1}{2\sigma^2_e} \left( \sum_{j=1}^M \sum_{j' \neq j}^{M} \mathbb{E}[\gamma_j \tilde{\beta}_j] \mathbb{E}[\gamma_{j'} \tilde{\beta}_{j'}] \bfx^T_j \bfx_{j'}\right) \\
&-\frac{M}{2} \log(2\pi \sigma^2_{\beta})  - \frac{1}{2\sigma^2_{\beta}} \sum_{j=1}^{M} \mathbb{E}[\tilde{\beta}_j^2] \\
 &+ \log \pi \sum_j \mathbb{E}[\gamma_j] + \log(1-\pi) \sum_j \mathbb{E}[1-\gamma_j] \\
      &+ \sum_{j=1}^M \pi_j \sum_{k=1}^K q_{11}^k (\log(\alpha_k p_{jk}^{\alpha_k-1} u_k)) + q_{10}^{k}\log(1-u_k)  \\
      &+ \sum_{j=1}^M (1-\pi_j) \sum_{k=1}^K q_{01}^k(\log(\alpha_k p_{jk}^{\alpha_k-1} (1-v_k)) + q_{00}^k \log(v_k).
\end{split}
\end{equation}
Because  $ q(\tilde{\bfbeta},\bfgamma, \bfGamma) =  q(\tilde{\bfbeta}|\bfgamma) q(\bfgamma) q(\bfGamma | \bfgamma)$,
the second term of Eq.~\eqref{eq:lq1} could be written as
\begin{equation}\label{eqlq2}
-\mathbb{E}_q[\log q(\tilde{\bfbeta},\bfgamma, \bfGamma)] =  -\mathbb{E}_q[\log q(\tilde{\bfbeta},\bfgamma)] -\mathbb{E}_q[\log q(\bfGamma | \bfgamma)].
\end{equation}

The two terms of Eq.~\eqref{eqlq2} could be evaluated respectively, we have
\begin{equation}
  \begin{split}
    -\mathbb{E}_q[\log q(\tilde{\bfbeta},\bfgamma)]  &= \frac{M}{2} \log \sigma^2_{\beta} + \sum_j \frac{1}{2}\pi_j(\log s^2_j - \log \sigma^2_{\beta}) \\
    & \quad - \sum_j(\pi_j \log \pi_j + (1-\pi_j)\log(1-\pi_j))  \\
  \end{split}
\end{equation}
and 
\begin{equation}
\begin{split}
  q(\bfGamma|\gamma) &=   \prod_{j=1}^M  \left( \prod_{k=1}^K \left(  \frac{\alpha_k p^{\alpha_k - 1}_{jk} u_k}{\alpha_k p^{\alpha_k - 1}_{jk} u_k + (1-u_k)} \right)^{\Gamma_{jk}} \left( \frac{1 - u_k}{\alpha_k p^{\alpha_k - 1}_{jk} u_k + (1-u_k)} \right)^{1 - \Gamma_{jk}} \right)^{\gamma_j} \\
         & \left( \prod_{k=1}^K \left( \frac{\alpha_k p^{\alpha_k - 1}_{jk} (1-v_k)}{\alpha_k p^{\alpha_k - 1}_{jk} (1-v_k) + v_k} \right)^{\Gamma_{jk}} \left( \frac{v_k}{\alpha_k p^{\alpha_k - 1}_{jk} (1-v_k) + v_k} \right)^{1 - \Gamma_{jk}} \right)^{1-\gamma_j} \\
         & =  \prod_{j=1}^M  \left( \prod_{k=1}^K (q^{jk}_{11})^{\Gamma_{jk}} (q^{jk}_{10})^{1 - \Gamma_{jk}} \right)^{\gamma_j} \left( \prod_{k=1}^K (q^{jk}_{01})^{\Gamma_{jk}} (q^{jk}_{00})^{1 - \Gamma_{jk}} \right)^{1- \gamma_j}, \\
\end{split}
\end{equation}
then
\begin{equation}
  -\mathbb{E}_q [\log q(\bfGamma | \gamma) ]  = -\left( \sum_{j=1}^M \pi_j \left( \sum_{k=1}^K q_{11}^{jk} \log q^{jk}_{11} + q_{10}^{jk} \log q^{jk}_{10} \right) + (1-\pi_j) \left(  \sum_{k=1}^K q_{01}^{jk} \log q^{jk}_{01} + q_{00}^{jk} \log q^{jk}_{00} \right) \right ).
\end{equation}

Now we substitute $\mathbb{E}[\gamma_j \tilde{\beta}_j] = \pi_j \mu_j$, $\mathbb{E}[(\gamma_j \tilde{\beta}_j)^2] = \pi_j(s^2_j + \mu^2_j)$, $\mathbb{E}[\tilde{\beta}_j^2] =  \pi_j(s^2_j + \mu^2_j) + (1-\pi_j) \sigma^2_{\beta}$, $\mathbb{E}[\gamma_j] = \pi_j$ and $\mathbb{E}[1-\gamma_j] = 1-\pi_j.$

We rearrange the lower bound
\begin{equation}\label{eqlowerbound}
  \begin{split}
   \mathcal{L}(q) & = \mathbb{E}_q[\log \pr(\bfy,\bfP ,\tilde{\bfbeta},\bfgamma|\bfX,\theta)] -\mathbb{E}_q[\log q(\tilde{\bfbeta},\bfgamma,\bfGamma)] \\
   &= -\frac{N}{2} \log(2\pi \sigma^2_e) - \frac{\|\bfy - \sum_j \pi_j\mu_j \bfx_j\|^2}{2\sigma^2_e} - \frac{1}{2\sigma^2_e}\sum_{j=1}^M[\pi_j(s^2_j + \mu^2_j) - (\pi_j \mu_j)^2]\bfx^T_j \bfx_j \\
   &\quad  -\frac{M}{2} \log(2\mathrm{\pi}) - \frac{1}{2\sigma^2_{\beta}} \sum_{j=1}^{M}[\pi_j(\mu^2_j + s^2_j) +(1-\pi_j)\sigma^2_{\beta}] \\
   &\quad + \sum_j \pi_j \log(\frac{\pi}{\pi_j}) + \sum_{j} (1-\pi_j) \log(\frac{1-\pi}{1-\pi_j})  \\
   & \quad + \sum_j \frac{1}{2} \pi_j(\log s^2_j - \log \sigma^2_{\beta}) \\
   &+ \sum_{j=1}^M \pi_j \sum_{k=1}^K q_{11}^{jk} \log(\alpha_k p_{jk}^{\alpha_k-1} u_k) + q_{10}^{jk}\log(1-u_k)  \\
      &+ \sum_{j=1}^M (1-\pi_j) \sum_{k=1}^K q_{01}^{jk}(\log(\alpha_k p_{jk}^{\alpha_k-1} (1-v_k)) + q_{00}^{jk} \log(v_k) \\
      &-\left( \sum_{j=1}^M \pi_j \left( \sum_{k=1}^K q_{11}^{jk} \log q^{jk}_{11} + q_{10}^{jk} \log q^{jk}_{10} \right) + (1-\pi_j) \left(  \sum_{k=1}^K q_{01}^{jk} \log q^{jk}_{01} + q_{00}^{jk} \log q^{jk}_{00} \right) \right ), \\
  \end{split}
\end{equation}
simplifying the last three rows of Eq.~\eqref{eqlowerbound}, we could get the following equation,
\begin{equation}
  \begin{split}
  &\sum_{j=1}^M \pi_j \sum_{k=1}^K q_{11}^{jk} \log(\alpha_k p_{jk}^{\alpha_k-1} u_k) + q_{10}^{jk}\log(1-u_k)  \\
      &+ \sum_{j=1}^M (1-\pi_j) \sum_{k=1}^K q_{01}^{jk}(\log(\alpha_k p_{jk}^{\alpha_k-1} (1-v_k)) + q_{00}^{jk} \log(v_k) \\
      &-\left( \sum_{j=1}^M \pi_j \left( \sum_{k=1}^K q_{11}^{jk} \log q^{jk}_{11} + q_{10}^{jk} \log q^{jk}_{10} \right) + (1-\pi_j) \left(  \sum_{k=1}^K q_{01}^{jk} \log q^{jk}_{01} + q_{00}^{jk} \log q^{jk}_{00} \right) \right ) \\
       &= \sum_{j=1}^M \pi_j \sum_{k=1}^K  \log (\alpha_k p_{jk}^{\alpha_k-1} u_k + 1 - u_k) +   \sum_{j=1}^M (1-\pi_j) \sum_{k=1}^K  \log (\alpha_k p_{jk}^{\alpha_k-1} (1-v_k) + v_k).
  \end{split}
\end{equation}

To get $\pi_j$, we set $\frac{\partial \mathcal{L}(q)}{\partial \pi_j} = 0$, yielding
\begin{equation}
     \pi_j = \frac{1}{1 + \exp(-w_j)},\mathrm{where}~ w_j = \log \frac{\pi}{1-\pi} + \frac{1}{2} \log \frac{s^2_j}{\sigma^2_{\beta}} + \frac{\mu_j^2}{2s^2_j}   + \sum_{k=1}^K \log \frac{\alpha_k p_{jk}^{\alpha_k -1 } u_k + 1 - u_k}{\alpha_k p_{jk}^{\alpha_k -1 } (1 -v_k) + v_k}. 
\end{equation}

\subsubsection{M-Step}
We will update the model parameters $\bftheta = \{\sigma^2_{\beta},\sigma^2_{e},\pi, \{\alpha_k\}_{k=1}^K\}$ sequentially by maximizing the lower bound $\mathcal{L}(q)$. \\
To get $\sigma^2_{\beta}$, we set
\begin{equation}
  \begin{split}
    \frac{\partial \mathcal{L}(q)}{\partial \sigma^2_{\beta}} &= \frac{\partial \mathbb{E}_q[\log \pr(\bfy,\tilde{\bfbeta},\bfgamma,\bfP |\bfX,\bftheta)]}{\partial \sigma^2_{\beta}}  = 0, \\
  \end{split}
\end{equation}
yielding
\begin{equation}
  \sigma^2_{\beta} = \frac{\sum_j \pi_j (\mu^2_j + s^2_j)}{\sum_j \pi_j}.
\end{equation}

To get $\sigma^2_{e}$, we set
\begin{equation}
  \begin{split}
    \frac{\partial \mathcal{L}(q)}{\partial \sigma^2_{e}} = \frac{\partial \mathbb{E}_q[\log \pr(\bfy,\tilde{\bfbeta},\bfgamma,\bfP |\bfX,\bftheta)]}{\partial \sigma^2_{e}}  &= -\frac{N}{2} \frac{1}{\sigma^2_e} + \frac{\bfy^T \bfy}{2\sigma^4_e} \\
    & - \frac{\sum_j \mathbb{E}[\gamma_j \tilde{\beta}_j]\bfx^T_j \bfy}{\sigma^4_e} \\
    & + \frac{1}{2 \sigma^4_e} \sum_{j=1}^M (\mathbb{E}[(\gamma_j \tilde{\beta}_j)^2] \bfx^T_j \bfx_j )  \\
    & + \frac{1}{2 \sigma^4_e } \left( \sum_{j=1}^M \sum_{j' \neq j}^{M} \mathbb{E}[\gamma_j \tilde{\beta}_j] \mathbb{E}[\gamma_{j'} \tilde{\beta}_{j'}] \bfx^T_j \bfx_{j'}\right) = 0,
  \end{split}
\end{equation}
yielding
\begin{equation}
   \begin{split}
      \sigma^2_e &= \frac{1}{N}\left(  \bfy^T \bfy - 2\sum_{j} \mathbb{E}[\gamma_j\tilde{\beta}_j] \bfx^T_j \bfy  + \sum_{j=1}^M (\mathbb{E}[(\gamma_j \tilde{\beta}_j)^2] \bfx^T_j \bfx_j ) + \sum_{j=1}^M \sum_{j' \neq j}^{M} \mathbb{E}[\gamma_j \tilde{\beta}_j] \mathbb{E}[\gamma_{j'} \tilde{\beta}_{j'}] \bfx^T_j \bfx_{j'} \right)  \\
  &= \frac{1}{N}\left(  \bfy^T \bfy - 2\sum_{j} \mathbb{E}[\gamma_j\tilde{\beta}_j] \bfx^T_j \bfy  + \sum_{j=1}^M \sum_{j' = 1}^{M} \mathbb{E}[\gamma_j \tilde{\beta}_j] \mathbb{E}[\gamma_{j'} \tilde{\beta}_{j'}] \bfx^T_j \bfx_{j'} + \sum_{j=1}^M (\mathbb{E}[(\gamma_j \tilde{\beta}_j)^2] \bfx^T_j \bfx_j ) - \sum_{j=1}^M   (\mathbb{E}[\gamma_j \tilde{\beta}_j])^2 \bfx_j^T \bfx_j   \right)  \\
  &= \frac{1}{N} \left( \|\bfy - \sum_j \pi_j \mu_j \bfx_j\|^2 + \sum_{j=1}^M(\pi_j(s_j^2 + \mu_j^2)-(\pi_j \mu_j)^2)\bfx_j^T \bfx_j \right ).
   \end{split}
\end{equation}

To get $\pi$, we set $\frac{\partial \mathcal{L}(q)}{\partial \pi} = 0$, yielding
\begin{equation}
  \pi = \frac{1}{M} \sum_j \pi_j.
\end{equation}

The parameter of the beta distribution, $\alpha_k$, could be obtained by maximizing $\mathcal{L}(q)$,
\begin{equation}
  \alpha_k = \frac{\displaystyle \sum_{j=1}^{M} \pi_j q^{jk}_{11} + (1-\pi_j)q^{jk}_{01}}{ \displaystyle  \sum_{j=1}^{M} \left( \pi_j q^{jk}_{11} + (1-\pi_j)q^{jk}_{01} \right) (-\log(p_{jk}))}.
\end{equation}

The parameters $u_k$ and $v_k$ could also be obtained by maximizing $\mathcal{L}(q)$, we have
\begin{equation}
  u_k = \frac{\sum_{j=1}^M \pi_j q^{jk}_{11}}{\sum_{j=1}^M \pi_j},
\end{equation}
\begin{equation}
  v_k = \frac{\sum_{j=1}^M (1-\pi_j)q^{jk}_{00} }{\sum_{j=1}^M (1-\pi_j)}.
\end{equation}

\subsection{ALGORITHMS}
\subsubsection{Basic Algorithm Steps}

Now we describe an  algorithm:
 \begin{itemize}
   \item $\mathrm{Initialize}$ $\{\pi_j, \mu_j \}_{j=1}^M, \sigma^2_{\beta}, \sigma^2_e, \{\alpha_k\}_{k=1}^K, \{q_{00}^{jk}\}, \{q_{11}^{jk}\}$. Let $\tilde{\bfy} = \sum_j \pi_j \mu_j \bfx_j.$
   \item $\mathrm{E-Step}$: For $j=1,\hdots,M$, first obtain 
\begin{equation}{\label{algorithmeq1}}
    \tilde{\bfy}_j = \tilde{\bfy} - \pi_j \mu_j \bfx_j,
\end{equation}

   and then update $\mu_j, s^2_j,\pi_j$ and $\tilde{\bfy}$ as follows
     \begin{align}
      \label{aeq1}      s^2_j &= \frac{\sigma^2_e}{\bfx^T_j \bfx_j + \sigma^2_e / \sigma^2_{\beta} }, \\
     \label{aeq2}           \mu_j &= \frac{\bfx^T_j(\bfy - \tilde{\bfy}_j)}{\bfx^T_j \bfx_j + \sigma^2_e / \sigma^2_{\beta} }, \\
     \label{aeq3}           \pi_j &= \frac{1}{1 + \exp(-w_j)},\mathrm{where}~   w_j = \log \frac{\pi}{1-\pi} + \frac{1}{2} \log \frac{s^2_j}{\sigma^2_{\beta}} + \frac{\mu_j^2}{2s^2_j}
          + \sum_{k=1}^K \log \frac{\alpha_k p_{jk}^{\alpha_k -1 } u_k + 1 - u_k}{\alpha_k p_{jk}^{\alpha_k -1 } (1 -v_k) + v_k},  \\
     \label{aeq4}           \tilde{\bfy} &= \tilde{\bfy}_j + \pi_j \mu_j \bfx_j.
     \end{align}
   \item $\mathrm{M-Step}$
     \begin{equation}
        \begin{split}
            \sigma^2_e &= \left( \|\bfy - \tilde{\bfy}\|^2 + \sum_{j=1}^M(\pi_j(s_j^2 + \mu_j^2)-(\pi_j \mu_j)^2)\bfx_j^T \bfx_j \right ) / N,  \\
                \alpha_k &= \frac{\displaystyle \sum_{j=1}^{M} \pi_j q^{jk}_{11} + (1-\pi_j)q^{jk}_{01}}{ \displaystyle  \sum_{j=1}^{M} \left( \pi_j q^{jk}_{11} + (1-\pi_j)q^{jk}_{01} \right) (-\log(p_{jk}))},\\ 
            u_k &= \frac{\sum_{j=1}^M \pi_j q^{jk}_{11}}{\sum_{j=1}^M \pi_j}, \\
            v_k &= \frac{\sum_{j=1}^M (1-\pi_j)q^{jk}_{00} }{\sum_{j=1}^M (1-\pi_j)},\\
            \sigma^2_{\beta} &= \frac{\sum_j \pi_j(\mu_j^2 + s^2_j)}{\sum_j \pi_j}, \\
            \pi &= \frac{1}{M} \sum_j \alpha_j.  
        \end{split}
     \end{equation}
   \item Evaluate the lower bound $\mathcal{L}(q)$, check the stop criteria.
\end{itemize}

According to the definition, we could update $q_{11}^{jk},q_{00}^{jk}$ by
\begin{equation}
    \begin{split}
    q^{jk}_{11} &= \frac{\alpha_k p^{\alpha_k - 1}_{jk} u_k}{\alpha_k p^{\alpha_k - 1}_{jk} u_k + (1-u_k)},  \\
    q^{jk}_{00} &= \frac{v_k}{\alpha_k p^{\alpha_k - 1}_{jk} (1-v_k) + v_k}.  
    \end{split}
\end{equation}






\subsection{Condition of irrelevance between the individual-level data and summary-level data}
If the association status for the individual-level data and the $k$-th trait are irrelevant, the following equation holds,
\begin{equation}\label{eq_independent}
    \Pr(\gamma_j =1 , \Gamma_{jk} = 1) = \Pr(\gamma_j=1)\Pr(\Gamma_{jk}=1).
\end{equation}
As defined in the main text, $s=\pr(\gamma_j = 1), u_k = \Pr(\Gamma_{jk} = 1 | \gamma_j =1)$, the left side of Eq.\eqref{eq_independent} could be written as
\begin{equation}
    \Pr(\gamma_j =1 , \Gamma_{jk} = 1) =  \Pr(\Gamma_{jk} = 1 | \gamma_j =1) \pr(\gamma_j =1) = u_k s.
\end{equation}
As
\begin{equation}
  \begin{split}
      \pr(\Gamma_{jk} = 1) &= \pr (\Gamma_{jk}=1|\gamma_j=1) \pr(\gamma_j=1) +  \pr (\Gamma_{jk}=1|\gamma_j=0) \pr(\gamma_j=0) \\
      &=u_k s + (1-v_k)(1-s),
  \end{split}
\end{equation}
When Eq.\eqref{eq_independent} holds, we have 
\begin{equation}
    u_k s = s (u_k s + (1-v_k)(1-s)).
\end{equation}
We could easily get the irrelevant condition for the $k$-th trait with Trait T as
\begin{equation}
    u_k + v_k = 1.
\end{equation}

\subsection{MORE RESULTS IN THE SIMULATION EXPERIMENTS}
In this section, we present more simulation results to the demonstrate the effectiveness of LEP. The results contains four parts, (1) Performance of FDR control under different pleiotropy settings; (2) Performance of FDR control with different number of summary-level datasets; (3) Performance of risk variant identification and risk prediction under homogeneous assumption; (4) The results for the case-control studies. As in the main text, the number of samples and the number of variants were set to be $N=2,000$ and $M=10,000$, respectively. The heritability  was pre-specified at 0.5 with the sparsity-level $s=0.05$.

\subsubsection{Performance of FDR control under different pleiotropy settings}
In the main text, we have compared LEP with BVSR, GPA and IGESS in terms of risk variants identification (AUC) and risk prediction under different pleiotropy settings (varying $u$ at {0.05, 0.3, 0.7, 0.9}), in this sub section, we list the results of FDR (False Discovery Rate) control. All the datasets in this sub section are simulated under different LD setting, $\rho = \{0,0.2,0.4,0.6,0.8\}$, and the results are based on 50 replications.

\begin{figure}[h!]
    \begin{center}
    \includegraphics[width=0.85\textwidth]{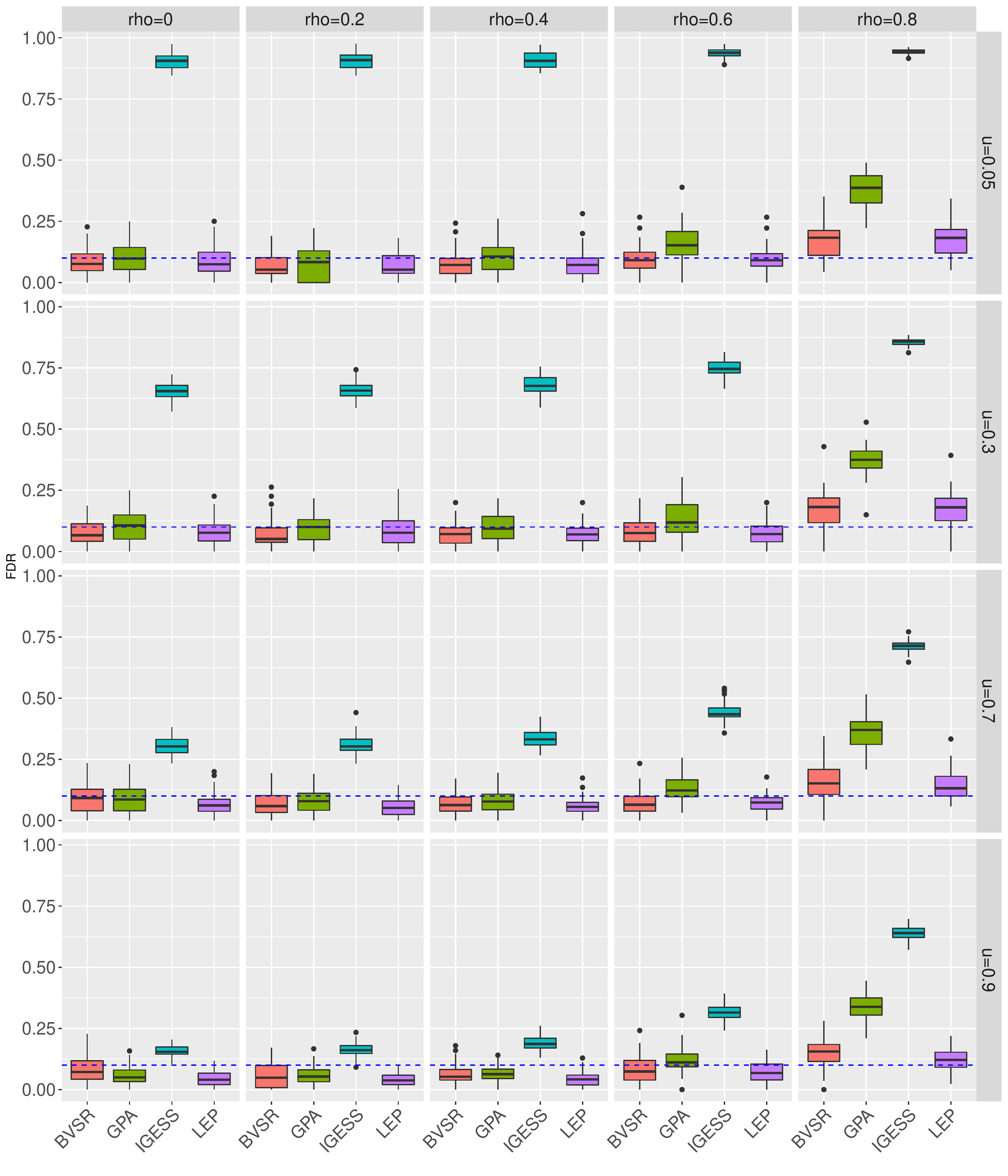}
    \centering
      \caption{Performance of False Discovery Rate (FDR) Control at pre-specified level FDR$=0.1$ under different pleiotropy settings and different LD. $K = 1$.}
          \label{AlongU_FDR}
    \end{center}
\end{figure}

As the results shown Fig.~\ref{AlongU_FDR}, LEP could control FDR well with small or moderate LD as $\rho = \{0.2,0.4,0.6\}$, we have observed slightly inflated FDR when $\rho = 0.8$.



\clearpage

\subsubsection{Performance of FDR control with different number of summary-level datasets}
In the main text, we have compared LEP with BVSR, GPA and IGESS in terms of variants risk identification(AUC) and risk prediction with different number of summary-level datasets, in this sub section, we evaluate the performance of FDR control.

\begin{figure}[h!]
    \begin{center}
    \includegraphics[width=0.80\textwidth]{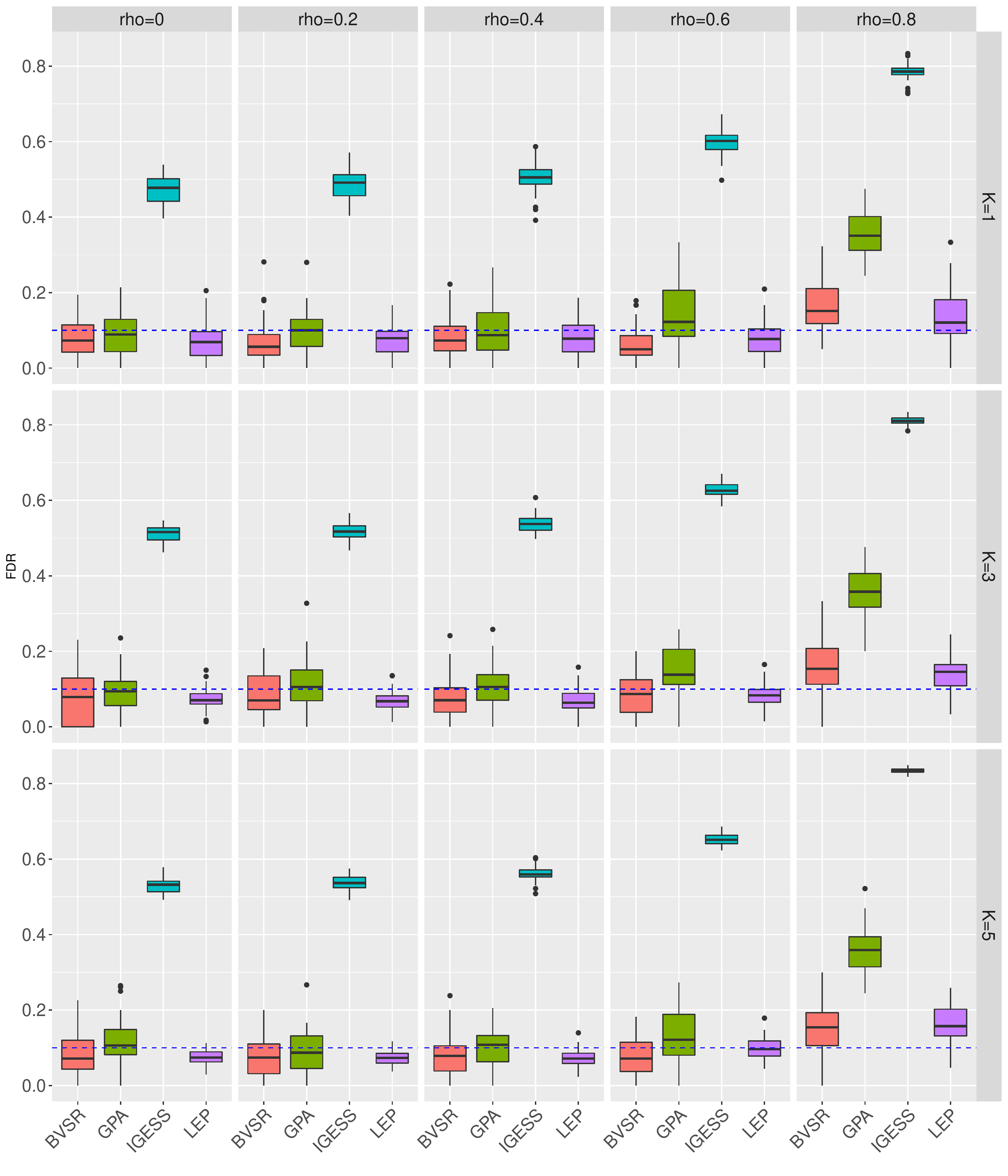}
      \caption{Performance of False Discovery Rate (FDR) Control at pre-specified level $FDR = 0.1$ under different pleiotropy settings and different LD. Here $u=0.5$}
          \label{BF_FDR}
    \end{center}
\end{figure}

According to the results in Fig.~\ref{BF_FDR}, LEP could also control the FDR at the pre-specified level FDR=$0.1$ when $\rho=\{0,0.2,0.4,0.6\}$. We have observed slightly inflated FDR when $\rho=0.8$.

\clearpage

\subsubsection{Results for the homogeneous assumption}
\begin{figure*}
  \begin{center}
  \includegraphics[width=0.85\textwidth]{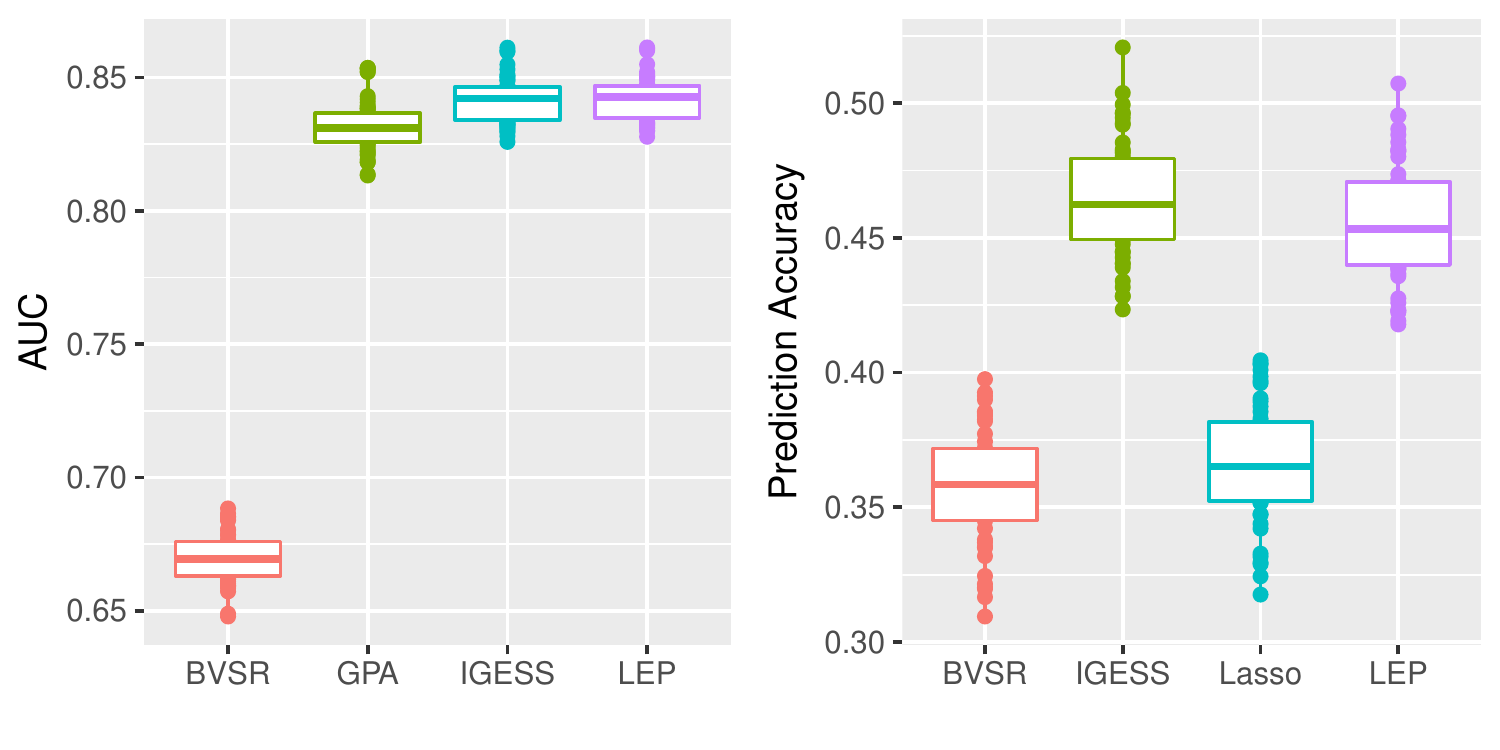}\quad
  \caption{Comparison of BVSR, GPA, IGESS and  Lasso. Performance of risk variant identification measured by AUC and Prediction Accuracy with a homogeneous assumption. The autoregressive correlation $\rho = 0$ and the number of summary statistics data sets $K=1$. All the results are summarized based on 50 replications.}
    \label{fig:aucPred}
  \end{center}
 \end{figure*}

We considered an extreme situation with a homogeneous assumption (the assumption of IGESS), $\Gamma_{j1} = \hdots = \Gamma_{jK}  = \bfGamma_j$, according to the results in Fig.~\ref{fig:aucPred}, we could see LEP has comparable performance with IGESS, LEP and IGESS all outperform BVSR and Lasso as they take in more information in the analysis effectively.

\clearpage

\subsubsection{Case-Control Studies}
\begin{figure}[h!]
    \begin{center}
    \includegraphics[width=0.75\textwidth]{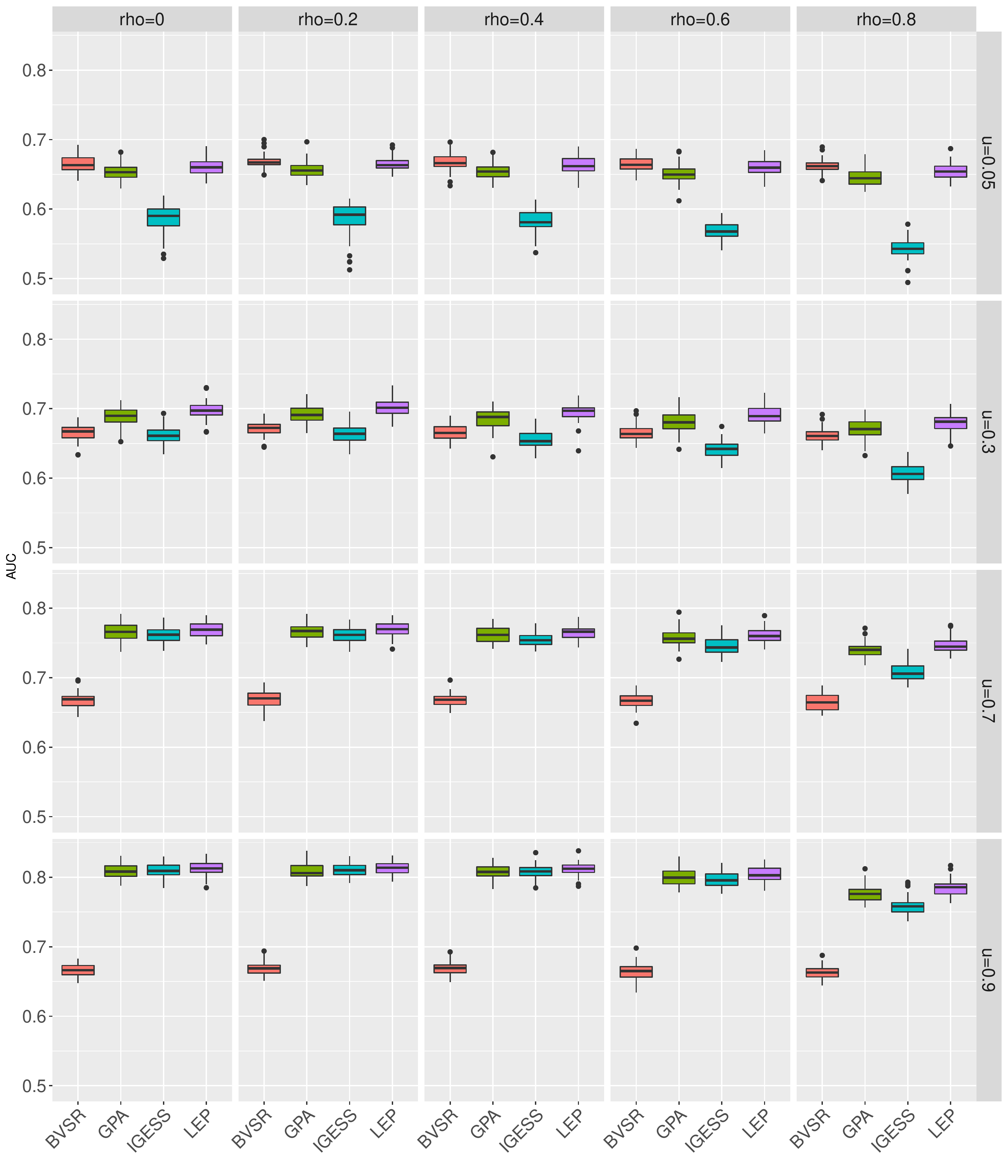}
    \caption{Performance of risk variants identification measured by AUC for the case-control studies}
        \label{case_control_AUC}
    \end{center}
\end{figure}

\begin{figure}[h!]
   \begin{center}
    \includegraphics[width=0.85\textwidth]{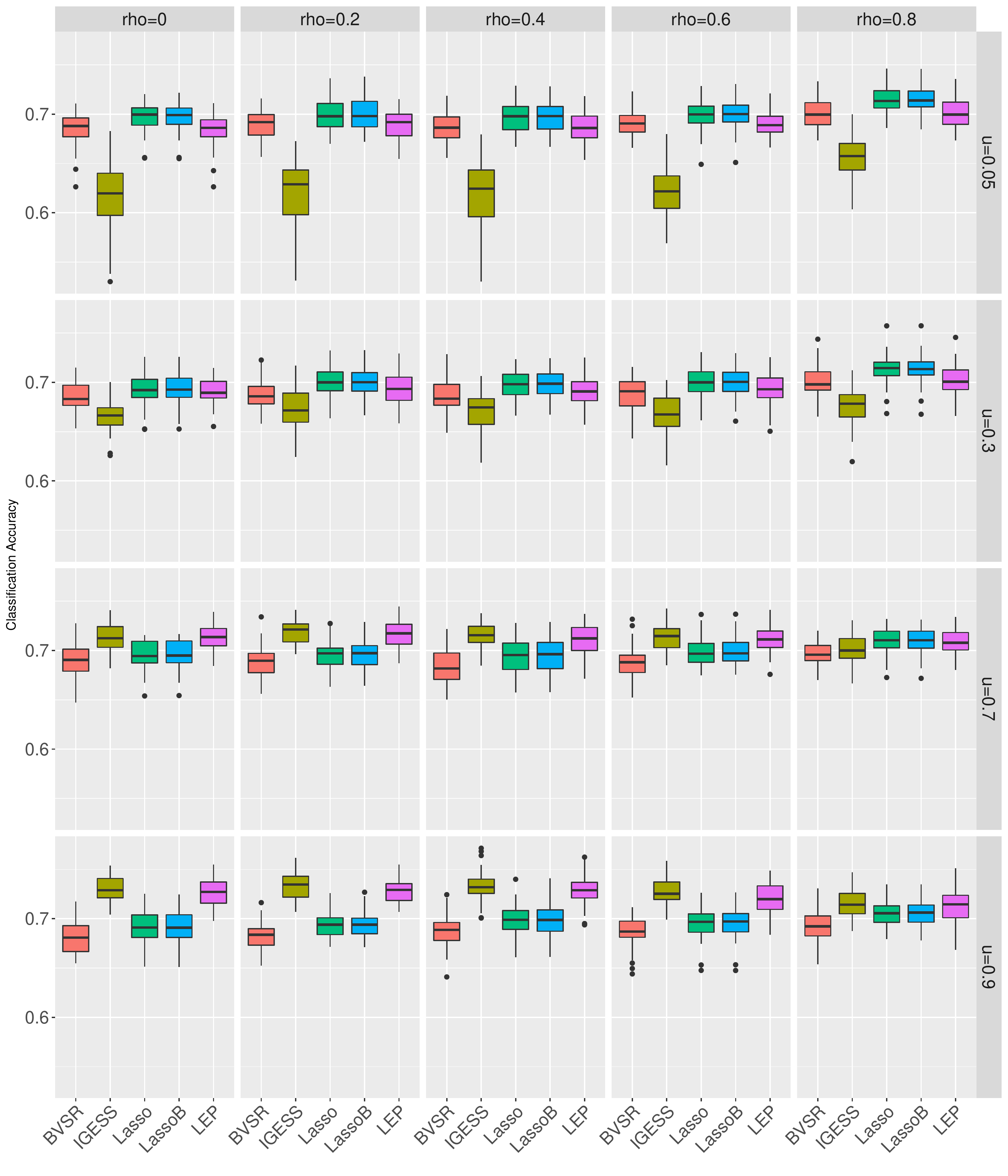}
    \caption{Performance of classification accuracy, which is measured by AUC evaluated on the observed phenotype values and the predicted values, Lasso corresponds to Lasso of Gaussian type, LassoB corresponds to Lasso of Binomial type. Lasso and LassoB have very similar performances.}
    \label{case_control_Predict}
    \end{center}
\end{figure}

\begin{figure}[h!]
    \begin{center}
    \includegraphics[width=0.85\textwidth]{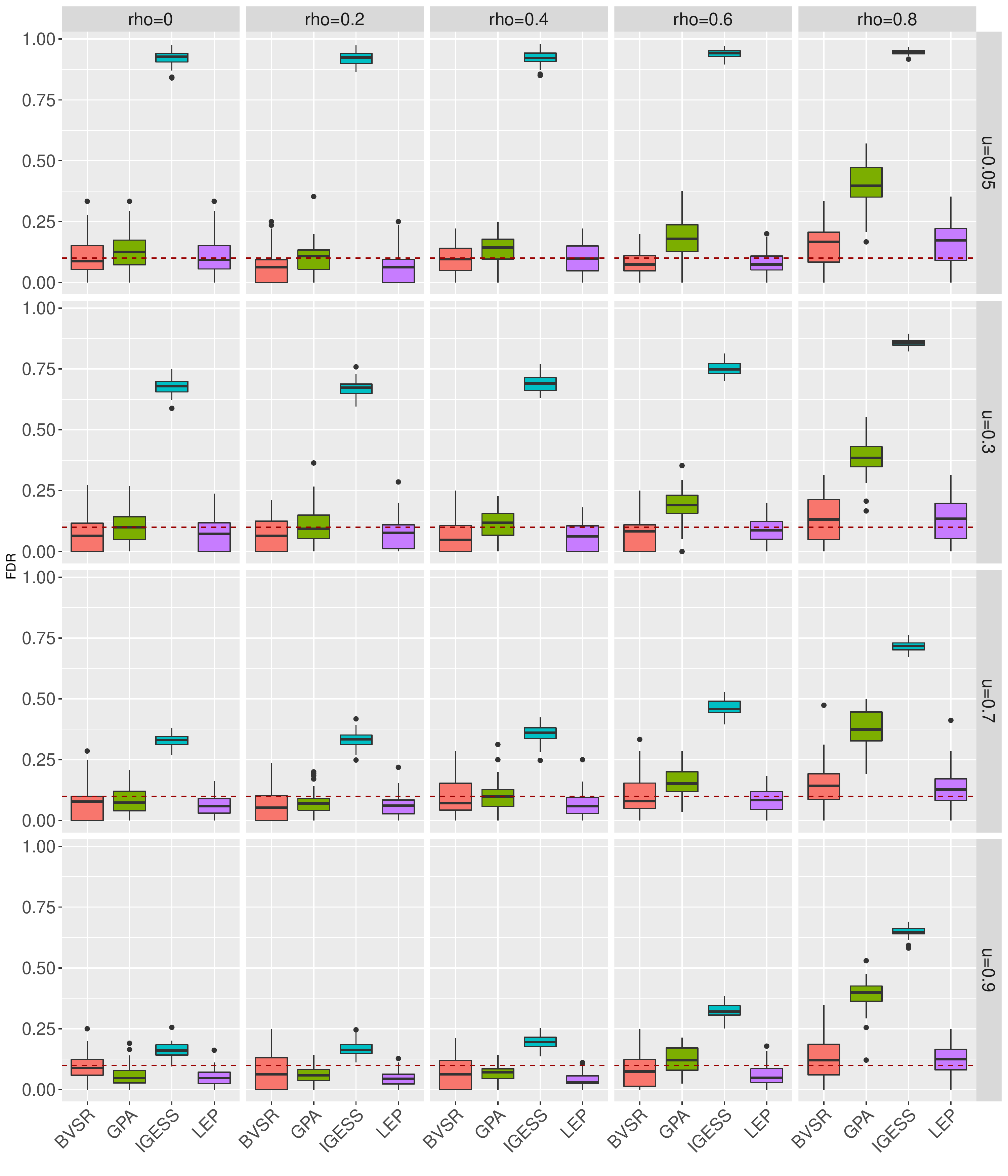}
    \caption{Performance of False Discovery Rate (FDR) Control at pre-specified level $FDR = 0.1$ for the case-control studies}
        \label{case_control_FDR}
    \end{center}
\end{figure}

In this sub section, we evaluate the performance of the case-control studies. Fig.~\ref{case_control_AUC} to Fig.~\ref{case_control_FDR} list the results of risk variants identification measured by AUC, classification accuracy, and False Discovery Rate (FDR) respectively. We fix $K=1$ and varies $u$ at $\{0.05, 0.3, 0.7, 0.9\}$ to mimic the different pleiotropy settings. All the results are summarized based on 50 replications. They take autoregressive correlation $\rho=\{0,0.2,0.4,0.6,0.8\}$ respectively. 

Fig.~\ref{case_control_AUC} indicate that LEP and GPA have comparable performances in terms of risk variants identification, they all outperform IGESS and BVSR.

\clearpage

Fig.~\ref{case_control_Predict} shows that LEP has comparable performance with BVSR where the summary-level data is irrelevant, LEP outperforms BVSR under all other settings which shows that the information from the summary-level data has been utilized effectively, and LEP outperforms Lasso and LassoB with low LD ($\rho = \{0,0.2,0.4,0.6\}$)  and higher pleiotropy effects ($u=\{0.7,0.9\}.$)


As the results shown Fig.~\ref{case_control_FDR}, LEP could control FDR well with small or moderate LD as $\rho = \{0.2,0.4,0.6\}$, we have observed slightly inflated FDR when $\rho = 0.8$ in the case-control studies.

\clearpage

\subsection{The source of the five GWAS of Summary Statistics}
We list the information in Table~\ref{t:basicinfo} for the summary statistics in the real data analysis, the information contains their associated sample size , their corresponding number of SNPs, their original paper and the link for downloading the datasets. 
\begin{table}[tp]
\centering
\begin{footnotesize}
\begin{tabular}{l|c|c|l}
\hline
   & N & M & Souce  \\
  \hline
  Alzheimer & 54,162 &1,136,071 &\tabincell{l}{ Lambert et al., 2013, Nature Genetics.  \\
  \url{https://data.broadinstitute.org/alkesgroup/sumstats_formatted/} } \\
  \hline
Neuroticism & 170,911 & 1,115,394& \tabincell{l}{Okbay et al., 2016a, Nature Genetics. \\ \url{http://ssgac.org/documents/Neuroticism_Full.txt.gz} } \\ 
  \hline
Type 1 Diabetes & 26,890 & 915,518& \tabincell{l}{Bradfield et al., 2011, PLoS Genetics. \\ \url{https://www.immunobase.org/downloads/protected_data/GWAS_Data} } \\ 
    \hline
  Ulcerative Colitis & 27,432 &1,076,835 &  \tabincell{l}{Jostins et al., 2012, Nature. \\ \url{https://data.broadinstitute.org/alkesgroup/sumstats_formatted/}} \\ 
   \hline
     Primary Biliary Cirrhosis &13,239 & 525,775 & \tabincell{l}{Cordell et al., 2015, Nature Communications \\ \url{https://www.immunobase.org/downloads/protected_data/GWAS_Data/} }\\ 
   \hline
\end{tabular}
\caption{Basic information for the summary statistics for Alzheimer, Neuroticism, Type 1 diabetes, Ulcerative Colitis, Primary Biliary Colitis} 
\label{t:basicinfo}
\end{footnotesize}
\end{table}

\paragraph{Funding}
This work was supported in part by grant NO. 61501389 and No. 11501440
from National Science Funding of China, grants NO. 12202114, NO. 22302815 and 
NO. 12316116 from the Hong Kong Research
Grant Council, and Initiation Grant NO. IGN17SC02 from University Grants Committee, startup grant R9405 from The Hong Kong University of Science and Technology, ITF391/15FX (P0162) from innovative technology funding of Hong Kong, Duke-NUS Medical School WBS: R-913-200-098-263, MOE2016-T2-2-029 from Ministry of Eduction, Singapore and Shenzhen Fundamental Research Fund under Grant No. KQTD2015033114415450.

\bibliography{referenceiLEP}

\bibliographystyle{natbib}
\bibliographystyle{achemnat}
\bibliographystyle{plainnat}
\bibliographystyle{abbrv}
\bibliographystyle{bioinformatics}

\bibliographystyle{plain}
%

\end{document}